\documentclass{aa}  

\usepackage{graphicx}
\usepackage{txfonts}
\usepackage{float}
\usepackage{color}
\begin{document}

   \title{Linearly Polarized Millimeter and Submillimeter Continuum Emission of Sgr A* Constrained by ALMA}


   \author{Hauyu Baobab Liu\inst{1}\fnmsep\inst{2}
          \and
          Melvyn C. H. Wright\inst{3}\fnmsep 
          \and
          Jun-Hui Zhao \inst{4}
          \and 
          Christiaan D. Brinkerink \inst{5}
          \and 
          Paul T.~P. Ho\inst{2}
          \and 
          Elisabeth A. C. Mills \inst{6}
          \and
          Sergio Mart\'in \inst{7}\fnmsep\inst{8}
          \and  
          Heino Falcke \inst{5}
          \and 
          Satoki Matsushita \inst{2}
          \and 
          Ivan Mart{\'{\i}}-Vidal \inst{9}
          }

    \institute{European Southern Observatory (ESO), Karl-Schwarzschild-Str. 2, D-85748 Garching, Germany \\
                  \email{baobabyoo@gmail.com}
          \and
              Academia Sinica Institute of Astronomy and Astrophysics, P.O. Box 23-141, Taipei, 106 Taiwan 
         \and
            Department of Astronomy, Campbell Hall, UC Berkeley, Berkeley, CA 94720
         \and
             Harvard-Smithsonian Center for Astrophysics, 60 Garden St, MS 78, Cambridge, MA 02138
         \and
         Department of Astrophysics/IMAPP Radboud University Nijmegen P.O. Box 9010 6500 GL Nijmegen The Netherlands
         \and 
         National Radio Astronomy Observatory, 1003 Lopezville Rd, Socorro, NM 87801, USA
         \and
         European Southern Observatory, Alonso de C\'ordova 3107, Vitacura, Santiago
         \and
            Joint ALMA Observatory, Alonso de C\'ordova 3107, Vitacura, Santiago, Chile
         \and 
          Onsala Space Observatory (Chalmers University of Technology), SE-43992 Onsala, Sweden
             }

   \date{Received April 15, 2016; accepted April xx, 20xx}


 
  \abstract
   {}
   {Our aim is to characterize the polarized continuum emission properties including intensity, polarization position angle, and polarization percentage  of Sgr A* at $\sim$100 (3.0 mm), $\sim$230 (1.3 mm), $\sim$345 (0.87 mm), $\sim$500 (0.6 mm), and $\sim$700 GHz (0.43 mm).}
   {We report continuum emission properties of Sgr A* at the above frequency bands, based on the Atacama Large Millimeter Array (ALMA) observations. We measured flux densities of Sgr A* from ALMA single pointing and mosaic observations. We performed sinusoidal fittings to the observed (XX-YY)/I intensity ratios, to derive the polarization position angles and polarization percentages.}
   {We successfully detect polarized continuum emission from all observed frequency bands. We observed lower Stokes I intensity at $\sim$700 GHz than that at $\sim$500 GHz, which suggests that emission at $\gtrsim$500 GHz is from optically thin part of a synchrotron emission spectrum. Both the Stokes I intensity and the polarization position angle at our highest observing frequency of $\sim$700 GHz, may be varying with time. However, we do not yet detect variation in the polarization percentage at $>$500 GHz. The polarization percentage at $\sim$700 GHz is likely lower than that at $\sim$500 GHz. By comparing the $\sim$500 GHz and $\sim$700 GHz observations with the observations at lower frequency bands, we suggest that the intrinsic polarization position angle of Sgr A* is varying with time. This paper also reports the measurable polarization properties from the observed calibration quasars.}
   {The future simultaneous multi-frequency polarization observations are required for clarifying the time and frequency variation of polarization position angle and polarization percentage.}


   \maketitle
%

\section{Introduction}
High angular resolution radio and (sub-)millimeter observations may trace relativistic accretion flows, or the footpoint of a jet, which are immediately around the Galactic supermassive black hole, Sgr A* (Falcke et al. 2000; Fish et al. 2009; Bower et al. 2014).
Modeling frameworks (see Yuan \& Narayan 2014 for a review of existing theories) with ray tracing calculations have suggested that observations of the linear polarization and time variability can constrain the morphology of the emitting gas and thereby the origin of the relativistic electrons (Bromley et al. 2001; Zhao et al. 2003, 2004; Liu et al. 2007; Marrone et al. 2006a, 2007; Falcke et al. 2009; Huang et al. 2009; Bower et al.2015; Brinkerink et al. 2015; Johnson et al. 2015; Liu et al. 2016).
In addition, the linearly polarized synchrotron emission from the relativistic electrons can be Faraday rotated by the accreting gas in the foreground, which can be diagnosed by multi-frequency observations of the linear polarization position angles and polarization percentages (Flett et al. 1991; Bower et al. 1999a, 1999c, 2001; Aitken et al. 2000; Bower et al. 2003, 2005; Macquart et al. 2006; Marrone et al. 2006a, 2007; Liu et al. 2016).
Moreover, a small fraction of linear polarization may be converted to circular, if the magnetized foreground screen is inhomogeneous and anisotropic, which can be tested by observing circular polarization (Bower et al. 1999b; Bower et al. 2002; Sault \& Macquart 1999; Mu{\~n}oz et al. 2012).
The degree of Faraday rotation by the foreground, and its time variability, will provide information about the black hole accretion rate and its time variability, which may (or may not) be related to the flaring activities of the Sgr A* (Zhao et al. 2003, 2004; Marrone et al. 2006a; Ponti et al. 2010; Clavel et al. 2013; Bower et al. 2015).
Kuo et al. (2014) for the first time derived the accretion rate of the active galactic nucleus of M87, based on the Submillimeter Array (SMA) observations of rotation measure.

In this work, we report new constraints on the polarized emission of Sgr A* at $\sim$90-710 GHz, based on Atacama Large Millimeter Array (ALMA) 12m-Array and Compact Array (ACA) single pointing and mosaic observations.
The 90-340 GHz observations we present are the first precisely constrained polarized emission properties within a single night, over such a wide range of observational frequencies.
In addition, our $\sim$700 GHz observations represent the highest frequency polarization measurements made by sub-millimeter interferometry so far, which provide valuable information from the optically thinnest part of the spectrum.
Details of our observations and data calibrations are provided in Section \ref{chap_obs}.
Our direct measurements are provided in Section \ref{chap_result}.
In Section \ref{chap_discussion} we address potential measurement errors.
We also compare our observational results with previous observations.
A brief conclusion is provided in Section \ref{chap_conclusion}.
We archive the fitting results of polarization percentage and polarization position angles for our calibrator quasars in Appendix \ref{appendix:cal}.


\begin{table*}
\caption{ALMA Observations}
\label{table:obs}
\hspace{0cm}
\centering
\begin{tabular}{ p{2.5cm} ccccc } \hline\hline
Frequency	&	UTC	&	Array	&	$uv$ distance range	& Flux/gain/passband calibrator & Gain calibrator flux\\
(GHz)		&		&			&	(meter)				&	& (Jy)				\\\hline
93/95/105/107 & 2012 May 18	& 12m	& 14-400	& Titan+Neptune/NRAO530/J1924-292	& 2.50 \\
245/247/261/263 & 2012 May 18	& 12m	& 14-400	& Titan+Neptune/NRAO530/J1924-292	& 1.45 \\
336/338/348/350 & 2012 May 18	& 12m	& 14-400	& Titan+Neptune/NRAO530/J1924-292	& 1.11 \\
480/482/490/492	&	2015 Apr. 30	& 12m	&	15-340	&  Titan/J1744-3116/J1833-2103 & 0.29 \\
480/482/490/492	&	2015 Apr. 30	& ACA	&	8.4-48	& Titan/J1744-3116/J1517-2422 & 0.30 \\
689/691/692/694 &	2015 May 02		& 12m	&	15-347	&	Titan/J1733-1304/J1924-2914	& 0.61 \\
689/691/692/694 &	2014 July 26		& ACA	&	8.5-49	&	Titan/J1733-1304/J1924-2914	& 0.55\\
706/708/710/712 &	2015 July 25-26		& ACA	&	8.2-86	&	Titan/J1733-1304/J1256-0547	& 0.95 \\
706/708/710/712 &	2015 July 26		& ACA	&	7.3-78	&	Titan/J1733-1304/J1256-0547	& 0.83 \\
683/684/686/688 &	2015 July 26		& ACA	&	8.5-89	&	Titan/J1733-1304/J1751+0939	& 0.82 \\\hline
\end{tabular}\par
\vspace{0.1cm}
\end{table*}

\section{Observations and Data Reduction} 
\label{chap_obs}
We analyzed the ALMA observations at band 3 ($\sim$3 mm), 6 ($\sim$1 mm), 7 ($\sim$0.88 mm), 8 ($\sim$0.6 mm), and 9 ($\sim$0.4 mm). 
These observations are summarized in Table \ref{table:obs}.
The receivers have orthogonal linearly polarized feeds. 
These observations measured the XX and YY linear correlations.
Except for band 7, the X polarization of the receivers is aligned radially in the receiver cryostat, with Y being aligned perpendicular to X.  
According to ALMA results from tests, the accuracy of this alignment is within 2 degrees.
The corresponding angular separations between the X polarization and the local vertical, which is known as {\it Evector}, is summarized in Table \ref{table:feed}.
Observations and data reduction for other frequency bands are introduced in the following Section \ref{sub:band36712m} and \ref{sub:band9}.
Our procedures to measure the intensities of the XX and the YY correlations, are introduced in Section \ref{sub:uvamp}.

\begin{figure}
\vspace{-2.0cm}
\hspace{-0.8cm}
\hspace{0.15cm}
\begin{tabular}{p{7.5cm} }
\includegraphics[width=10.5cm]{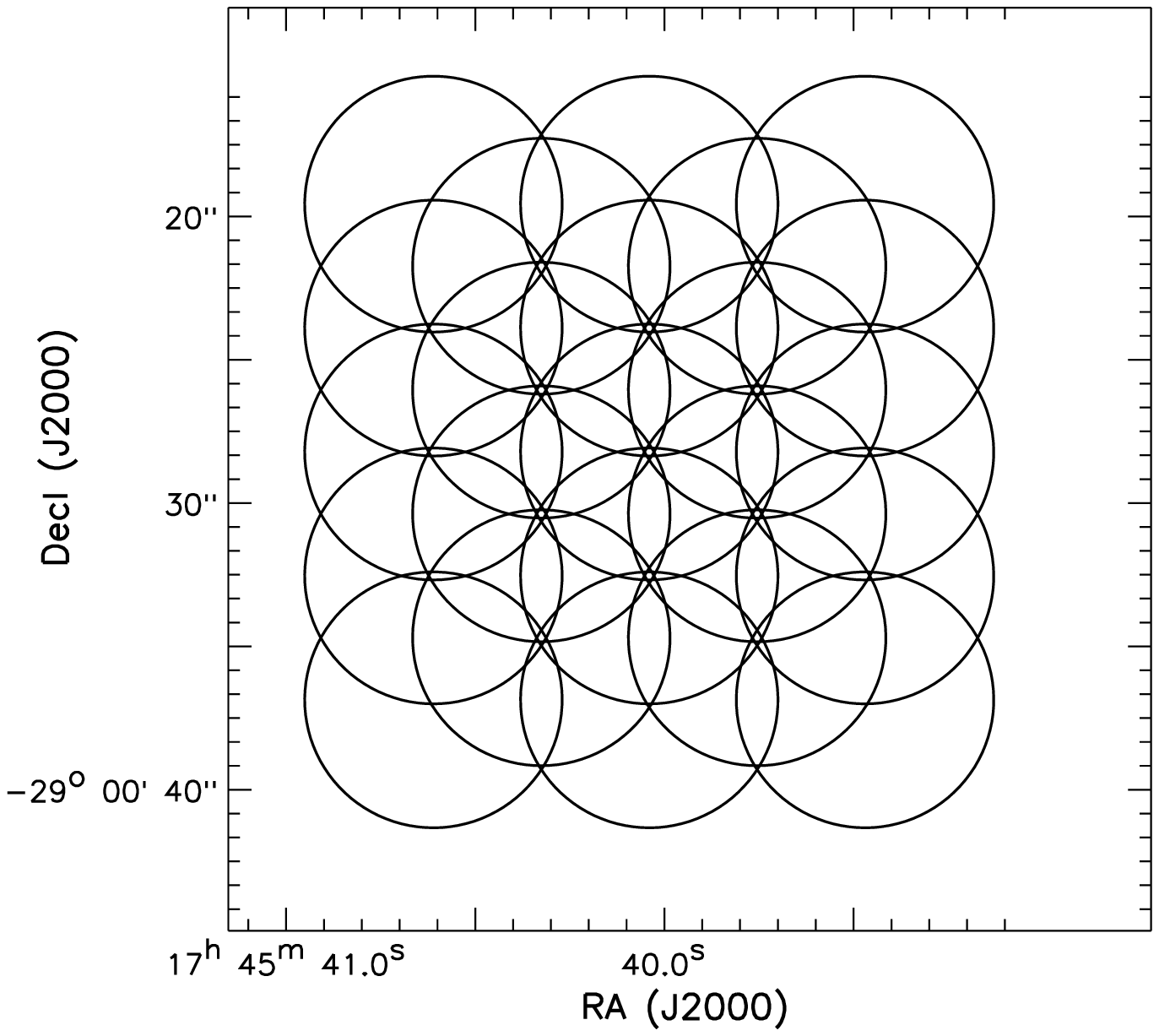} \\
\end{tabular}
\caption{\small{The mosaic fields of the ALMA 12m-Array, band 9 observations.
The diameter of these circles is $8\farcs9$.
}}
\label{fig:field12m}
\vspace{-0.1cm}
\end{figure}

\begin{figure}
\vspace{-2.0cm}
\hspace{-0.8cm}
\hspace{0.15cm}
\begin{tabular}{p{7.5cm} }
\includegraphics[width=10.5cm]{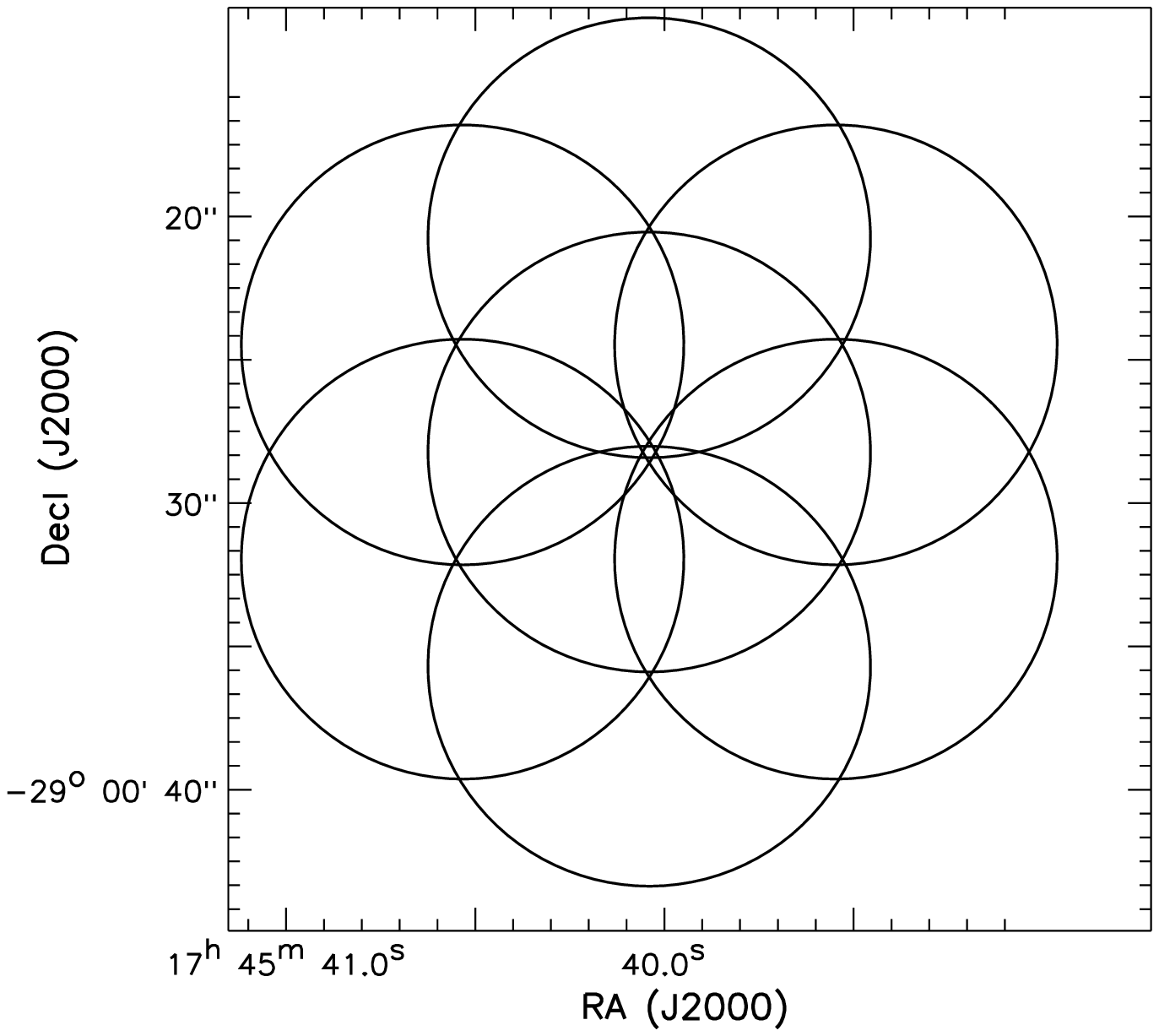} \\
\end{tabular}
\caption{\small{The mosaic fields of the ACA, band 9 observations.
The diameter of these circles is $15\farcs3$.
}}
\label{fig:fieldaca}
\vspace{-0.1cm}
\end{figure}

\subsection{ALMA 90-110, 250, 340, and 490 GHz observations (band 3, 6, 7, and 8)}\label{sub:band36712m}
The ALMA 12-m Array observations of 90-100 GHz (band 3), 250 GHz (band 6), and 340 GHz (band 7), were taken on UTC 2012 May 18 (UTC 03:30:47--10:52:16).
There were 19 available antennas for these observations.
This is a single-field observations with pointing and phase referencing centers approximately  coincide ($\sim$0$\farcs$01 offset) with the location of the Sgr A*.
The observations switched in between three frequency bands (3, 6 and 7); each frequency band was covered with a $\sim$7.5 GHz simultaneous bandwidth.
We referenced to the bright quasars J1924-2914 and J1733-1304 (also known as NRAO530) for passband and gain calibrations.
We refer to Brinkerink et al.(2015) for the calibration of the band 3, 6 and 7 observations.

The ALMA 12m Array and ACA mosaic observations of 490 GHz were taken on UTC 2015 April 30 (UTC 06:48:32--08:04:38).
The 7 mosaic fields of the 12m Array observations which covered the Sgr A*, are used for the analysis of the present paper.
There were 39 available antennas in the 12m Array observations.
The simultaneous bandwidth of these observations is $\sim$7.5 GHz.
We referenced to J1833-2103/J1517-2422 and J1733-3116 for passband and gain calibrations.
We refer the details of the band 8 observations to Liu et al. (2016).

\subsection{ALMA 680-710 GHz observations (band 9)}\label{sub:band9}
\subsubsection{12m array observations}\label{subsub:band912m}
The ALMA 12m-Array (consisting of 12 m dishes) mosaic observations of hexagonally packed 23 fields (Figure \ref{fig:field12m}) were carried out on 2015 May 02 (UTC 06:55:16.7--08:33:28), with 37 antennas.
These observations approximately covered a 25$''$$\times$25$''$ rectangular region.
The pointing and phase referencing center of the central field was R.A. (J2000) =17$^{\mbox{\scriptsize{h}}}$45$^{\mbox{\scriptsize{m}}}$40$^{\mbox{\scriptsize{s}}}$.036, and decl. (J2000) =-29$^{\circ}$00$'$28$''$.17, which is approximately centered upon Sgr A*.
We configured the correlator to provide four 1.875 GHz wide spectral windows (spws), covering the frequency ranges of 687.9-689.7 GHz (spw 0), 689.7-691.6 GHz (spw 1), 691.5-683.4 GHz (spw 2), and 693.3-695.2 GHz (spw 3), respectively. 
The frequency channel spacing was 3906.25 kHz.

The projected baseline length sampled by the 12m-Array observations is 15-347 meter.
The system temperature ($T_{sys}$) ranged from 600-1100 K.
We observed J1733-1304 approximately every 7.5 minutes for gain calibrations.
We observed Titan and J1924-2914 for absolute flux and passband calibrations, respectively.

\subsubsection{ACA observations}\label{subsub:band9aca}
The Atacama Compact Array (ACA) observations were performed with three different spectral setups:
\begin{description}
\item[B9-a:] 681.6-683.4 GHz (spw 0), 683.4-685.3 GHz (spw 1), 685.2-687.1 GHz (spw 2), and 687.0-688.9 GHz (spw 3)
\item[B9-b:] 705.2-707.0 GHz (spw 0), 707.0-708.8 GHz (spw 1), 708.8.5-710.6 GHz (spw 2), and 710.6-712.4 GHz (spw 3)
\item[B9-c:] 687.9-689.7 GHz (spw 0), 689.7-691.6 GHz (spw 1), 691.5-693.4 GHz (spw 2), and 693.3-695.2 GHz (spw 3)
\end{description}
All spectral setups were configured to provide 1.875 GHz wide spectral windows (spws) and a frequency channel spacing of 3906.25 kHz.
The B9-c setup is identical to the spectral setup of the 12m-array observations (Section \ref{subsub:band912m}).

We used hexagonally packed mosaic of 7 fields to cover Sgr A* approximately at the center (Figure \ref{fig:fieldaca}).
The ACA observations were carried out on 2014 July 26 (UTC 01:51:09.5--03:09:45.9; B9-c setup), 2015 July 25 (UTC 23:02:14.8-- July 26 01:01:17.2; B9-b setup), 2015 July 26 (UTC 01:16:20.3--03:08:00.6; B9-a setup), and 2015 July 26 (UTC 21:52:24.0--23:45:45.5; B9-b setup).
The ACA at the 2014 epoch consisted of ten dishes, which shared an identical (Mitsubishi, 7m) design. 
The ACA at the 2015 epochs additionally included three Mitsubishi 12m dishes to assist calibrations with the enhanced sensitivities to unresolved sources.

We excluded the ACA B9-c observations in our quantitative analysis, due to the relatively large amplitude errors.
The excluded ACA data were also the observations which were not assisted with the additional 12m dishes.
The $T_{sys}$ values of the rest of the ACA observations ranged from 500-1500 K.
The spatial samplings of the ACA observations (i.e. {\it uv} spacing range), and the calibrators we observed, are listed in Table \ref{table:obs}.
There are currently no available single-dish data to provide information on the zero-spacing fluxes for these observations.
However, the emission source of our interest is spatially very compact (Bower et al. 2014), and therefore we do not think missing short-spacing information is a concern for our measurements.

\begin{table}
\caption{Angular separation between X polarization and the local vertical (also known as Evector; c.f. Section 4.2 of ALMA Cycle 4 Technical Handbook)}
\label{table:feed}
\hspace{1.2cm}
\begin{tabular}{ c c c}\hline\hline
Band	& Frequency coverage &	Evector	 	\\
		&	(GHz)			&	(Degree)	\\\hline
3		& 84-116 &   -10.0   \\
6		& 211-275 &   -135.0 \\
7		& 275-373 &   -53.55 \\
8		& 385-500 &   0.0    \\
9		& 602-720 &   -180.0 \\\hline
\end{tabular}
\end{table}

\subsubsection{Data calibration}\label{subsub:band9reduction}
A priori calibrations including the application of $T_{sys}$ data, the water vapor radiometer (wvr) solution (which is only provided for the 12m-Array observations), antenna based passband calibrations, gain amplitude and phase calibrations, and absolute flux scaling, were carried out using the CASA software package (McMullin et al. 2007) version 4.5.
To enhance the signal to noise ratio, we first solved for and applied phase offsets between the four spectral windows, based on scans on the passband calibrator. 
We then derived gain calibration solutions. 
The gain phase solutions were derived separately for the XX and YY correlations, while the gain amplitude solutions were derived from the average of XX and YY correlations. 
We derived gain phase solutions averaging all spectral windows together. 
The calibration strategy for gain amplitude can avoid target source pick up the polarization properties of the gain calibrator.
This strategy valid when the gain amplitude variation is dominated by unpolarized effect (e.g. atmosphere).
The absolute flux scaling was derived incrementally from the gain amplitude solutions, combining all scans.

We fitted the continuum baselines from line-free channels, using the CASA task {\tt uvcontsub}.
After executing {\tt uvcontsub}, we generated a continuum dataset for each spectral window, by averaging the line free channels. 
We then exported the calibrated continuum data and the continuum-subtracted line data in standard fits format files, using the CASA task {\tt exportfits}.
Finally, we used the Miriad 4.3.8 (Sault et al. 1995) task {\tt fits} to convert the fits format data into the Miriad data format, for further analyses including imaging.

\subsection{Measuring fluxes}\label{sub:uvamp}
We modified the Miriad task {\tt uvamp}, to permit fitting fluxes for a point source as a function of the parallactic angle from the visibility data. 
We fit the fluxes of the XX and the YY correlations separately for every 5$^{\circ}$ bin of parallactic angle.
We found that this provides optimized signal to noise ratios, but without smearing too much of information in the parallactic angle (and time) domain.
For the band 3 data, we limited the fittings to data at $uv$ distance $>$30 $k\lambda$ (e.g. $\sim$7$''$ in terms of angular scale), to avoid the confusion from the bright and extended ionized mini-spiral arms (Lo \& Claussen 1983; Zhao et al. 2009, 2010).
The 12m antennas were omitted from the ACA before measuring Stokes I fluxes to avoid the potential bias between the absolute flux calibrations of the 7 and 12m antennas. 

For each epoch of observations, we derived the averaged Stokes I flux of Sgr A*.
For the band 3, 6, and 7 observations, the uncertainties of the Stokes I fluxes over a particular epoch of observations, were defined by the standard deviations of the Stokes I fluxes from all the 5$^{\circ}$ bins of parallactic angle.
For the band 8 and 9 observations, we assumed a nominal 20\% flux error, due to the relatively uncertain data calibrations.

\begin{table}
\caption{Stokes I fluxes of Sgr A*}
\label{table:flux}
\hspace{0.85cm}
\begin{tabular}{ c c c c}\hline\hline
Frequency	& Array & UTC	& Flux \\
(GHz)		&		&		& (Jy)	\\\hline
92.995		& 12m	& 2012May18	&  2.35$\pm$0.16 \\
94.932		& 12m	& 2012May18	&  2.37$\pm$0.16 \\
104.995		& 12m	& 2012May18	&  2.54$\pm$0.17 \\
106.995		& 12m	& 2012May18	&  2.62$\pm$0.18 \\
253.750		& 12m	& 2012May18	&  4.17$\pm$0.17 \\
342.979		& 12m	& 2012May18	&  4.26$\pm$0.24 \\
486.150		& 12m	& 2015Apr30	& 3.60$\pm$0.72	\\
691.537		& 12m	& 2015May02	& 2.68$\pm$0.54 \\
708.860		& ACA	& 2015Jul25/26 & 3.21$\pm$0.64 \\
685.237		& ACA	& 2015Jul26	& 2.65$\pm$0.53 \\
708.860		& ACA	& 2015Jul26	& 3.99$\pm$0.79 \\ \hline
\end{tabular}
\end{table}

\begin{figure}
\hspace{-1cm}
\includegraphics[width=10.5cm]{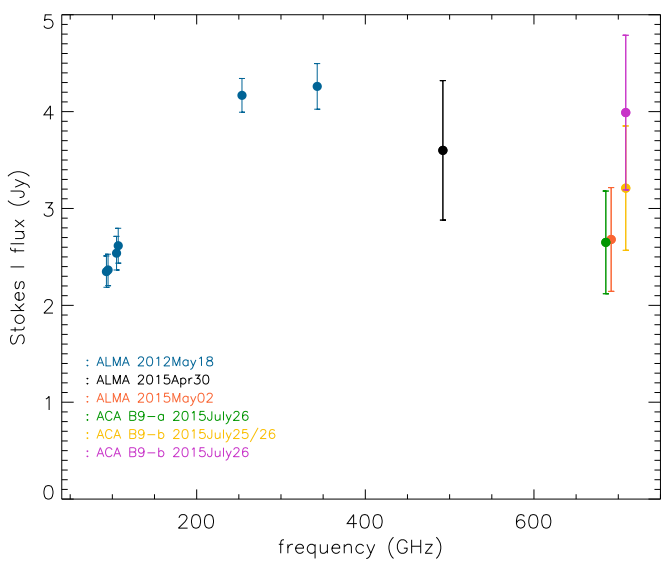}
\caption{Stokes I fluxes of the ALMA observations. Note the 3 year time interval between bands 3, 6, 7 and 8, 9. See Table \ref{table:obs} and Section \ref{chap_obs}. The vertical error bars are explained in Section \ref{sub:uvamp}.}
\label{fig:sed}
\end{figure}

\section{Results} 
\label{chap_result}
For the sake of conciseness, XX and YY will denote the fluxes (i.e. in Jy units) observed by these two correlations hereafter in this manuscript.
The Stokes I flux will simply be denoted by I.
The derived Stokes I fluxes will be summarized in Section \ref{sub:stokesI}.
Results of the Sgr A* polarization will be provided in Section \ref{sub:pol}.

\begin{figure*}
\begin{tabular}{p{9cm} p{9cm}}
\includegraphics[width=9.5cm]{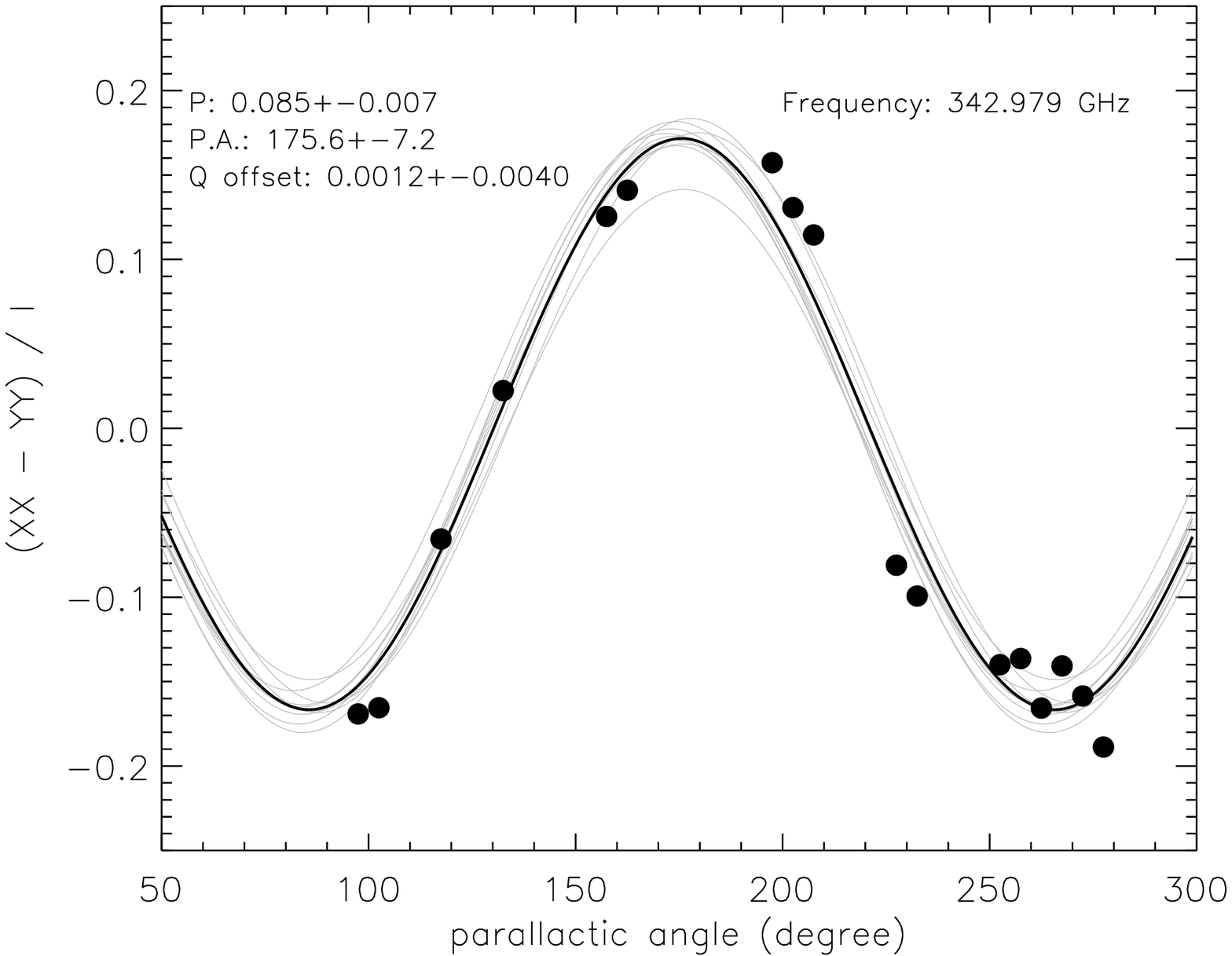} & \includegraphics[width=9.5cm]{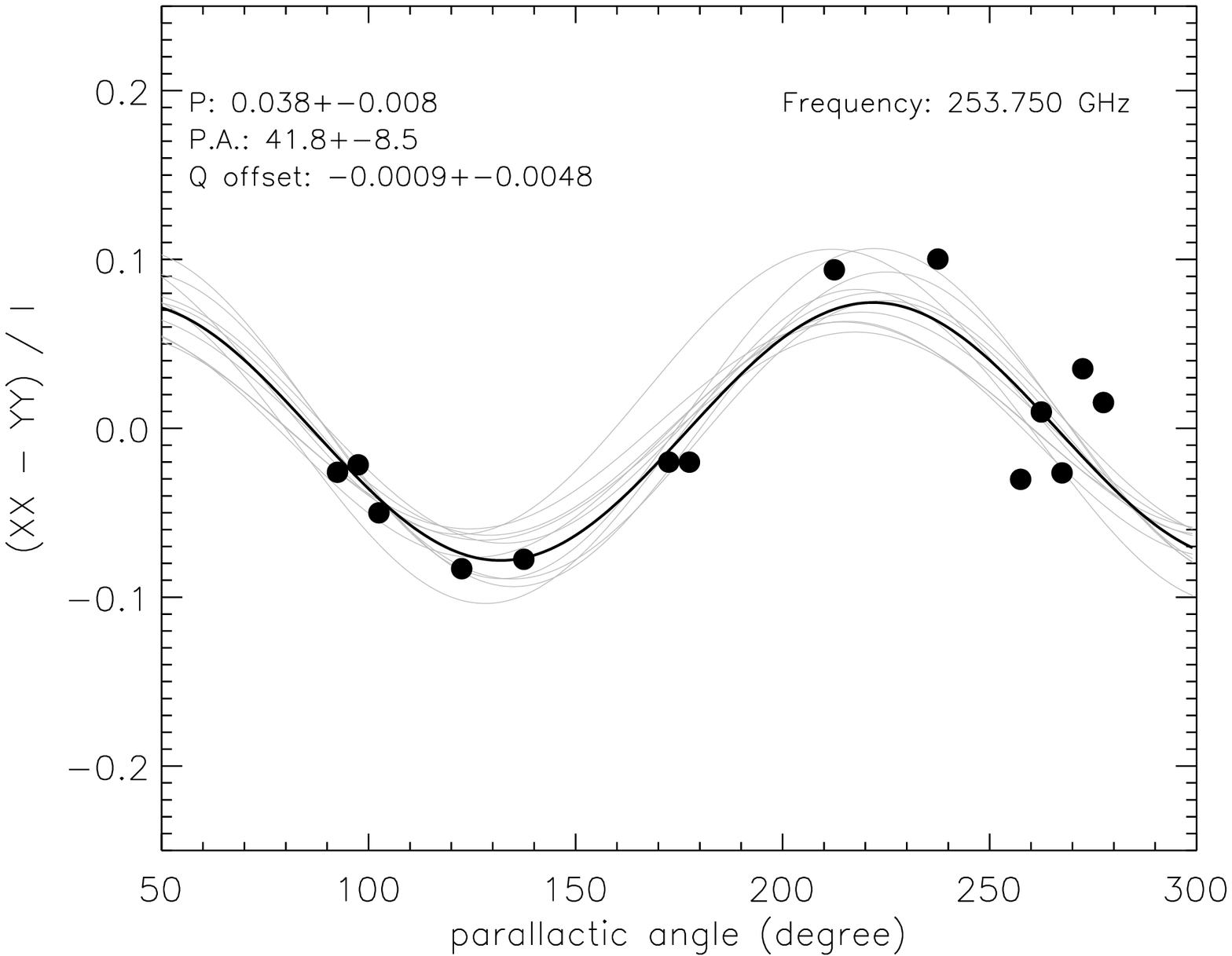} \\
\includegraphics[width=9.5cm]{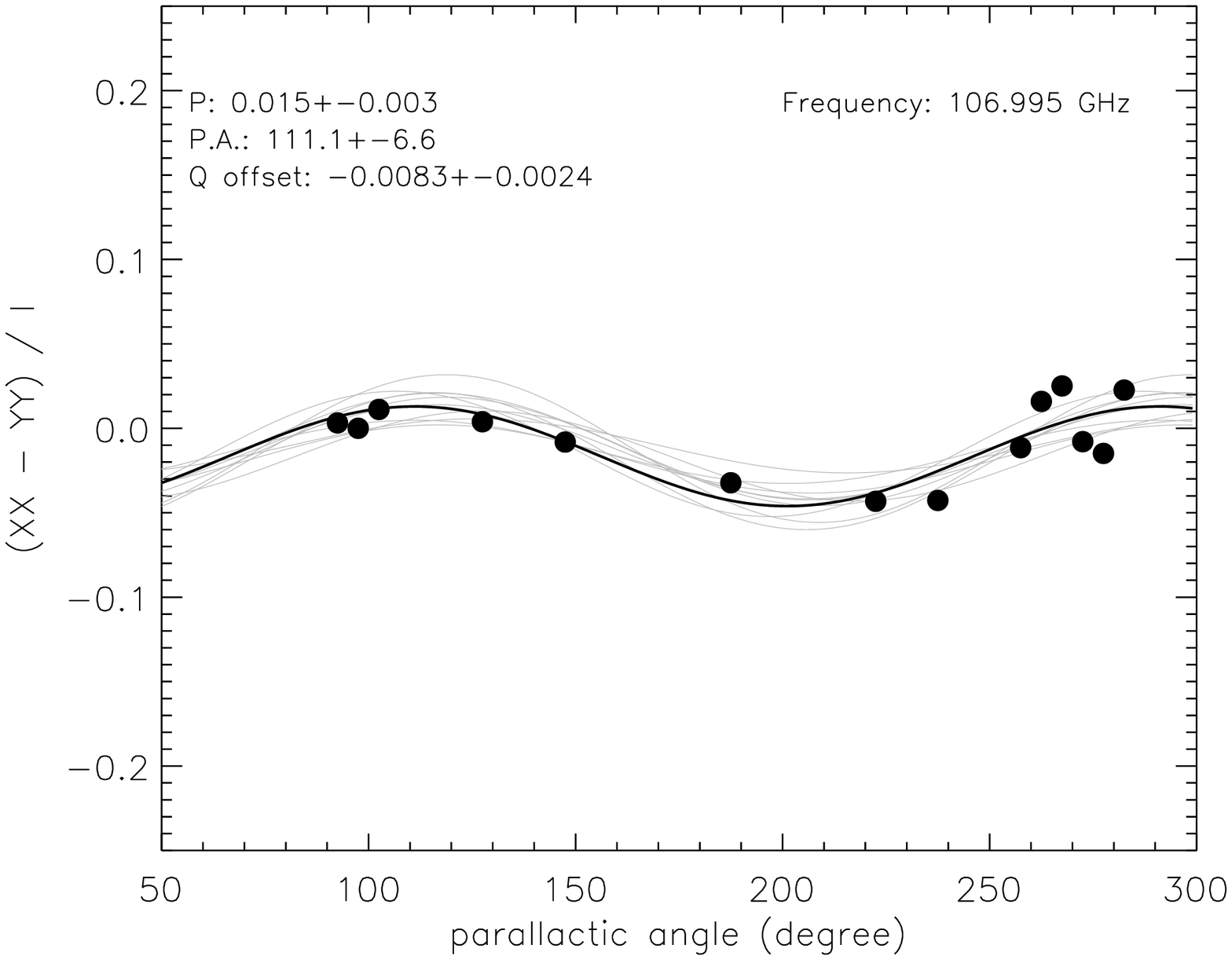} & \includegraphics[width=9.5cm]{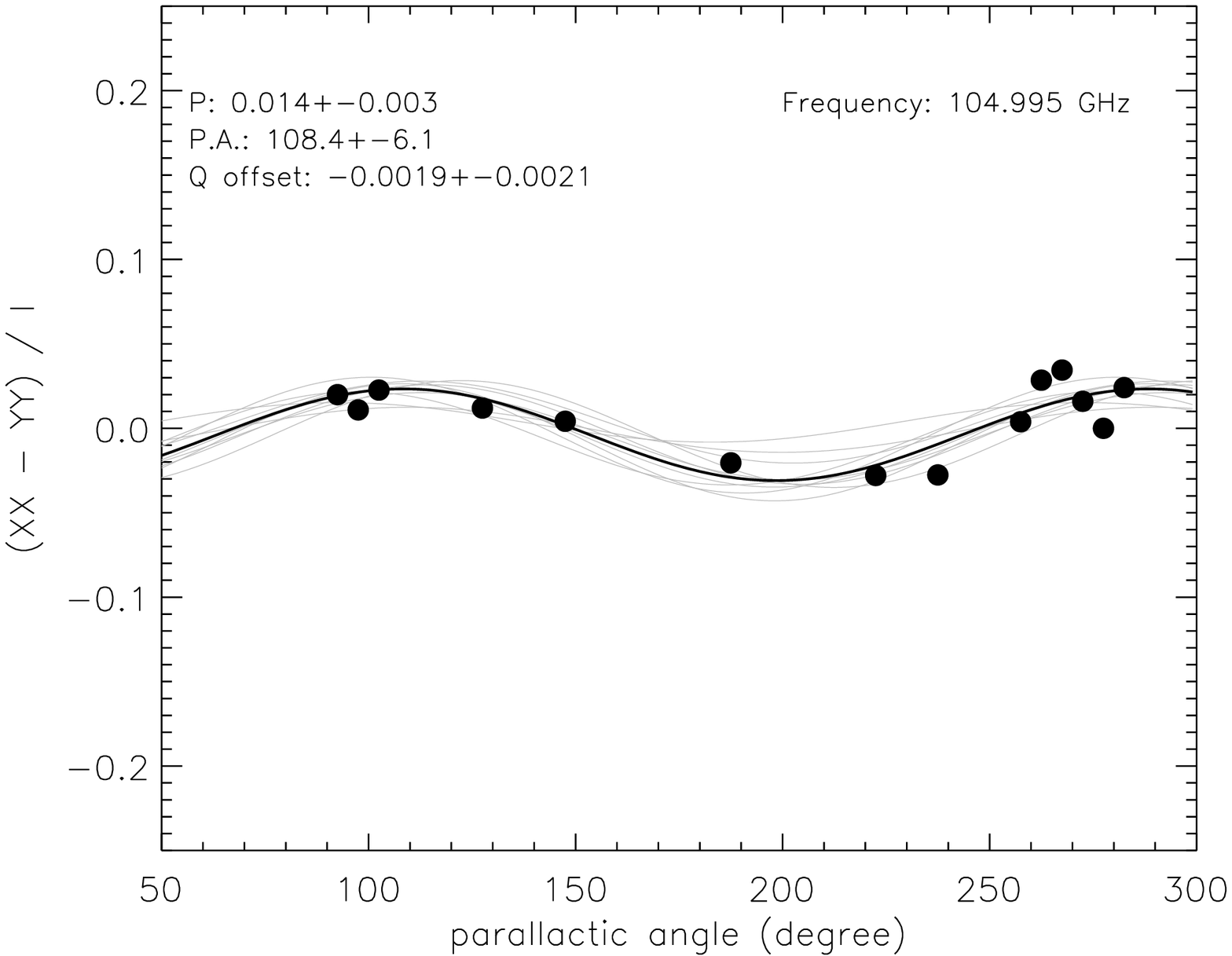} \\
\includegraphics[width=9.5cm]{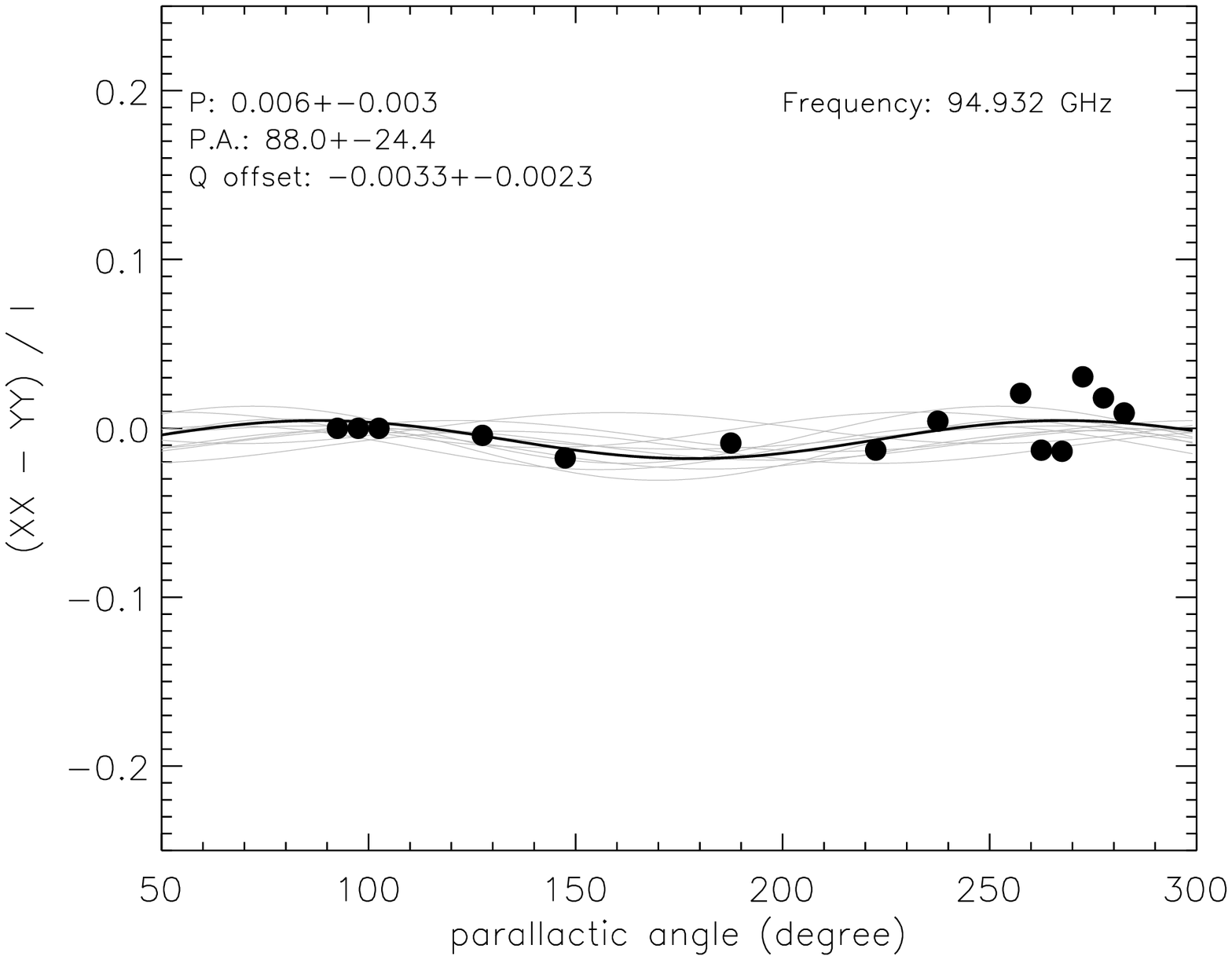} & \includegraphics[width=9.5cm]{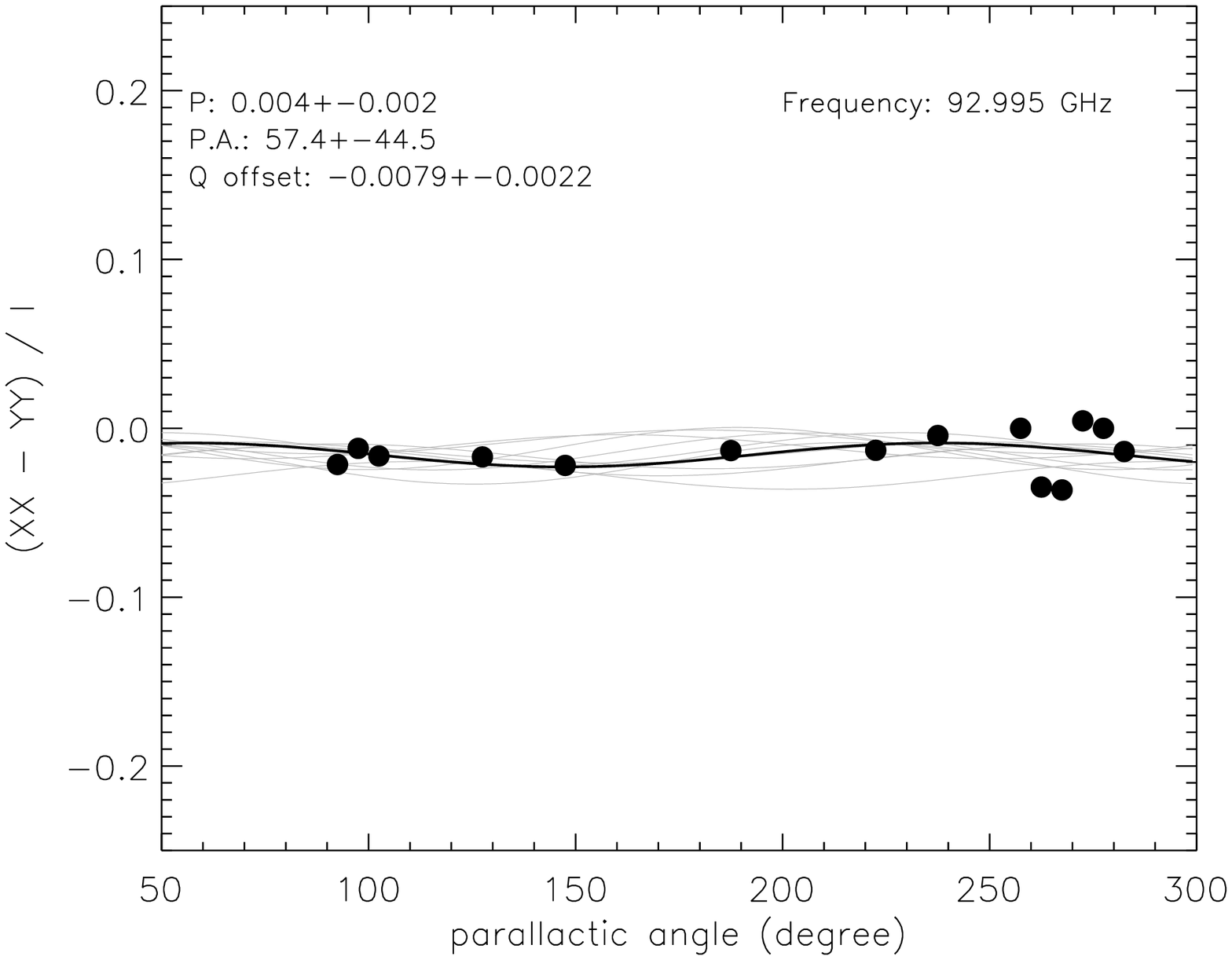} \\
\end{tabular}
\caption{Fittings of the (XX-YY)/I intensity ratio, to determine polarization percentages and polarization position angles. Data presented in this figure are self-calibrated ALMA band 3 (bottom 4 panels, from each of the four spectral windows of band 3), 6 (top right), 7 (top left) observations of the Sgr A* on UTC 2012 May 18. XX and YY are intensities of the two orthogonal polarizations in the receiver frame. Only the $uv$-sampling range of $>$30 $k\lambda$ was fitted for the band 3 (93-107 GHz) data. The best fits of polarization percentage (P), polarization position angle (in the receiver frame, i.e. $\Psi-\phi$; P.A.), and a constant normalized Stokes Q offsets (Q offset), are provided in the upper left of each panel, which are represented by a black curve. For each observed frequency, errors of fitted quantities were determined by one standard deviation of fittings of 1000 random realizations of noisy data (details are in Section \ref{sub:pol}). Gray lines in each panel plot every 100 of the random realizations.}
\label{fig:fit2012may18}
\end{figure*}

\begin{figure*}
\begin{tabular}{p{9cm} p{9cm}}
\includegraphics[width=9.5cm]{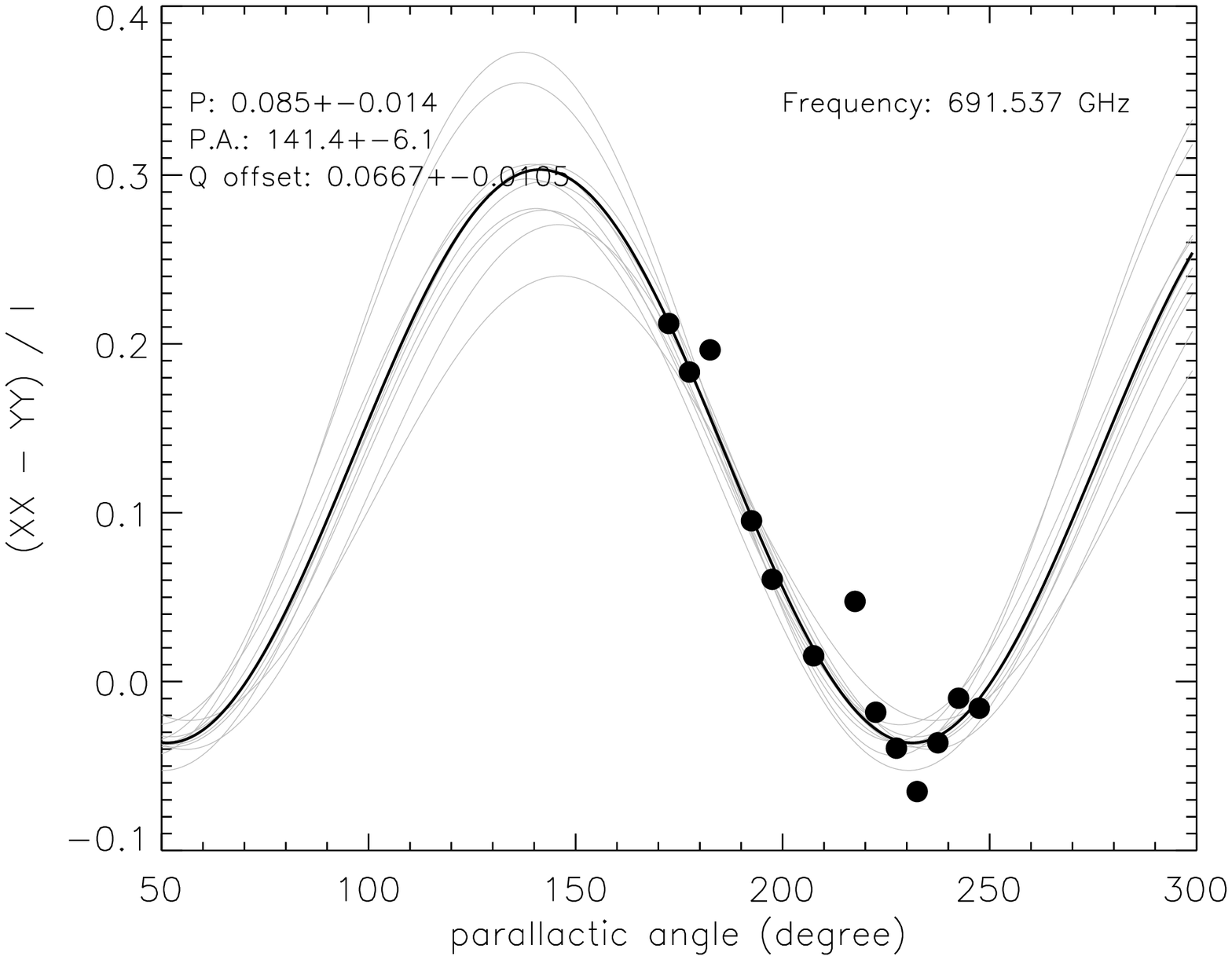} & \includegraphics[width=9.5cm]{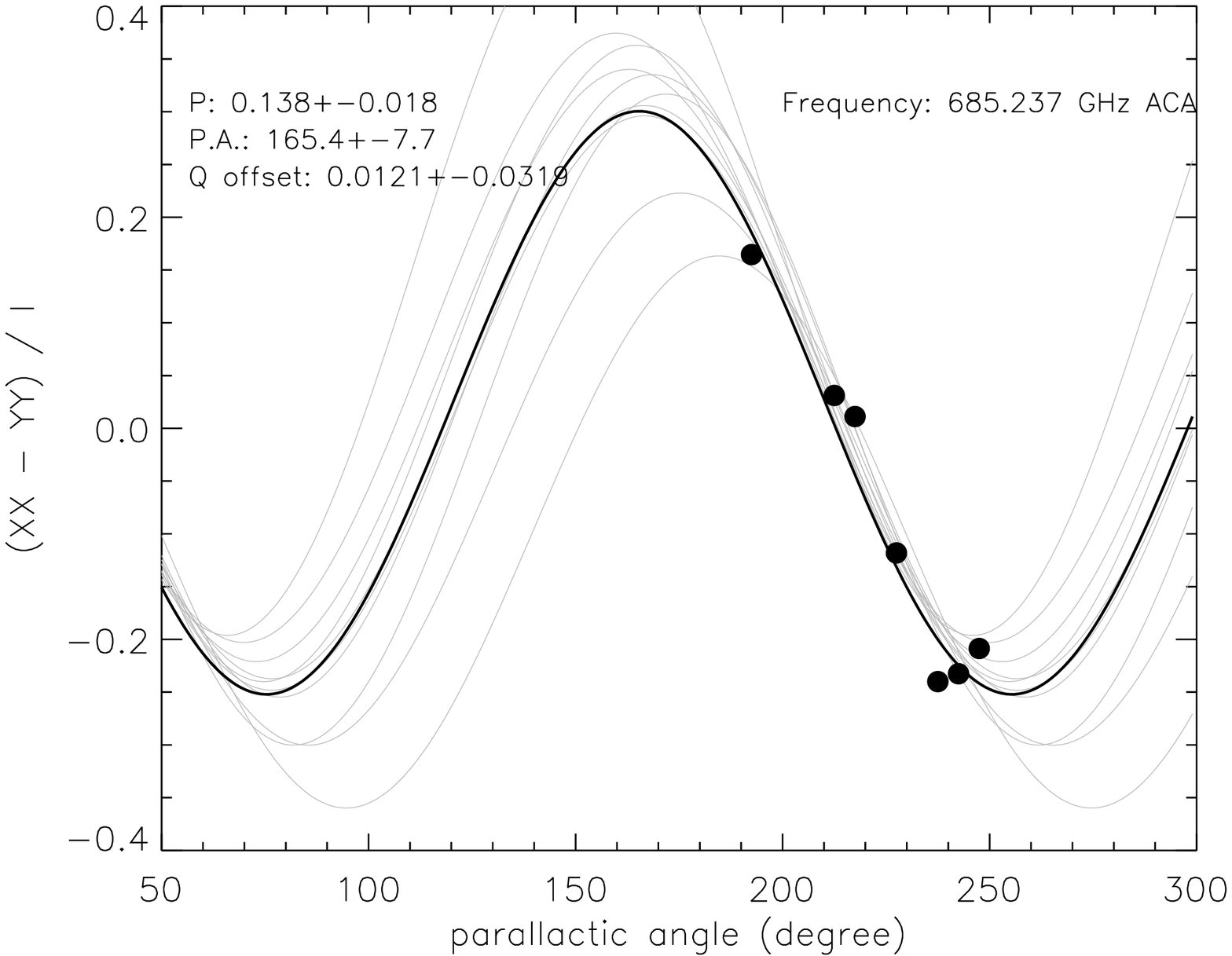} \\
\includegraphics[width=9.5cm]{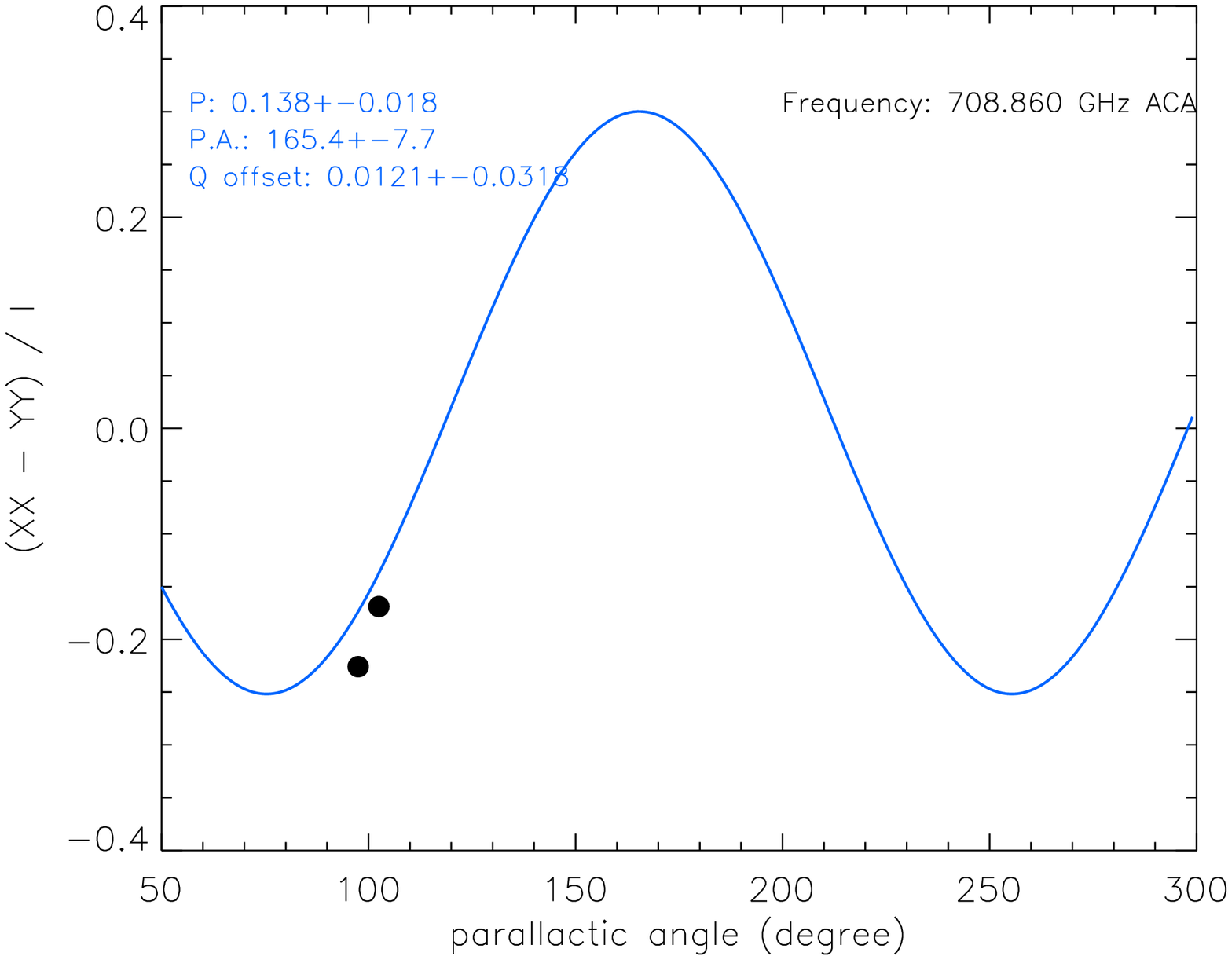} & \includegraphics[width=9.5cm]{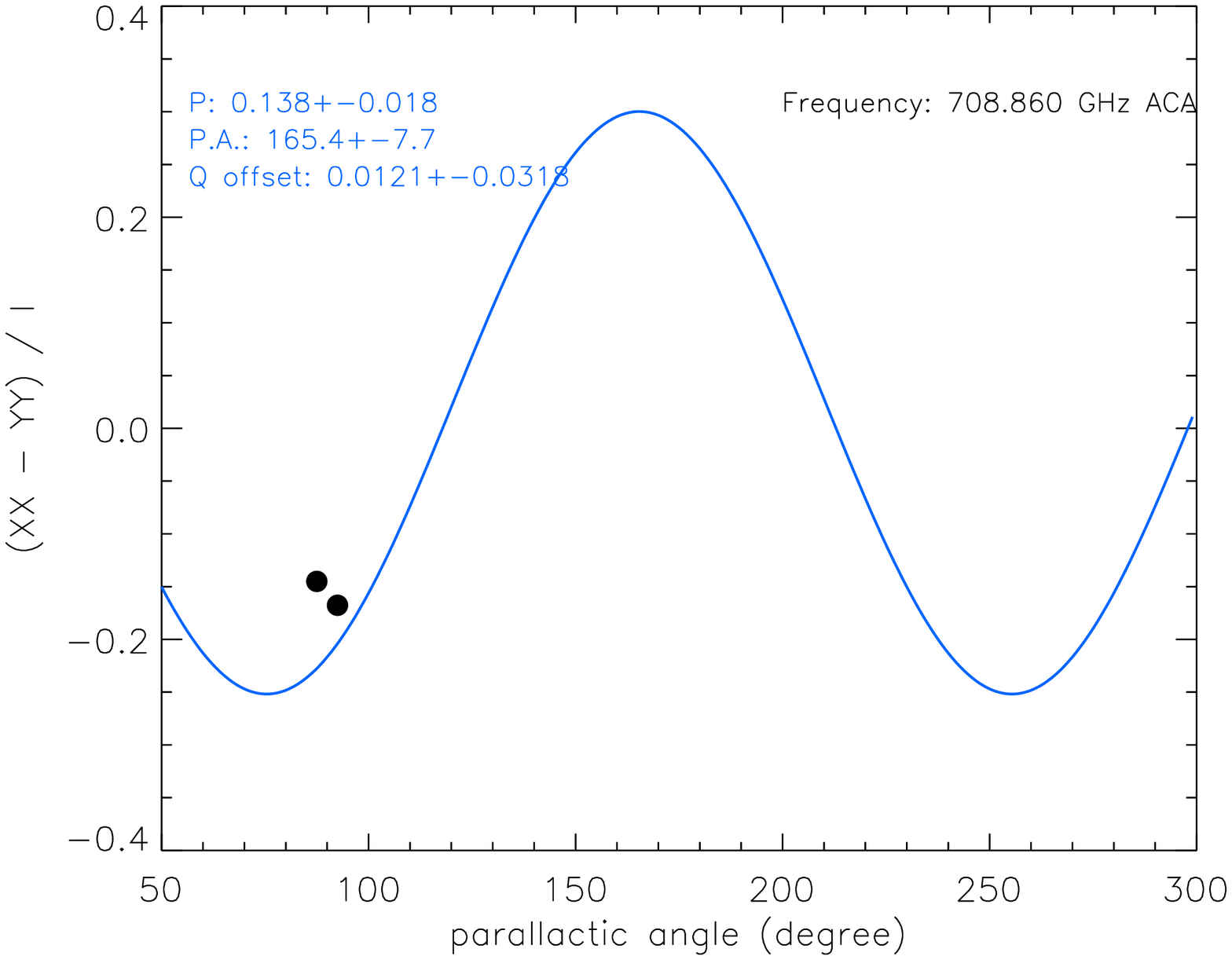}  \\
\end{tabular}
\caption{Fittings of the (XX-YY)/I intensity ratio, to determine polarization percentages and polarization position angles. Data presented in this figure are self-calibrated ALMA observations of the Sgr A* at band 9. XX and YY are intensities of the two orthogonal polarizations in the receiver frame. 
{\it Top left:--} The ALMA 12m-Array observations at passband setup-c (Table \ref{table:obs}).
{\it Top right:--} The ACA observations at passband setup B9-a.
{\it Bottom left:--} The ACA observations at passband setup B9-b on UTC 2015 July 25/26.
{\it Bottom right:--} The ACA observations at passband setup B9-b on UTC 2015 July 26.
The best fits of polarization percentage (P), polarization position angle (in the receiver frame, i.e. $\Psi-\phi$; P.A.), and a constant normalized Stokes Q offsets (Q offset), are provided in the upper left of each panel, which are represented by a black curve. For each observed frequency, errors of fitted quantities were determined by one standard deviation of fittings of 1000 random realizations of noisy data (details are in Section \ref{sub:pol}). Gray lines in each panel plot every 100 of the random realizations.
The ACA observations at setup B9-b only covered a small parallactic range, and therefore cannot provide independent constraints on P, P. A., and Q offset. 
Nevertheless, the observing UTC time and observing frequency of the two epochs of setup B9-b observations, are close to those of the setup B9-a observations, so may serve as consistency checks.
In blue color we overplot the best fit of the setup B9-a observations of ACA, to the panel which presents the setup B9-b data (bottom left and right panel).
}
\label{fig:fit2015b9}
\end{figure*}

\subsection{Stokes I flux density and spectral indices}
\label{sub:stokesI}
Figure \ref{fig:sed} shows the observed Stokes I fluxes of the Sgr A*, from individual epochs of ALMA observations, which were also summarized in  Table \ref{table:flux}.

Observations at $<$400 GHz, which were taken within a $\sim$7 hours period (Brinkerink et al. 2015), show increasing Stokes I fluxes with frequency.
However, the Stokes I fluxes at $\sim$250 GHz and $\sim$340 GHz are consistent within 1$\sigma$ error.
The $\sim$690 GHz observations on 2015 May 02 shows lower Stokes I flux than that of the $\sim$490 GHz observations taken on 2015 April 30.
If the apparently decreasing flux with observing frequency is not due to flux variability on the daily timescale (e.g. Dexter et al. 2014), then it may indicate that the $>$500 GHz observations are already in the optically thin regime (see also Falcke et al. 1998). 
This is also consistent with the data presented in An et al. (2005) and Doi et al. (2011).
If this is indeed the case, then the submillimeter very long baseline interferometry (VLBI) observations (e.g. Johnson et al. 2015, and references therein) at these frequencies, may be the key to penetrate to the innermost part of the gas accretion flow surrounding the Sgr A*.
However, the 709 GHz Stokes I fluxes obtained from the 2015 July 25-26 observations marginally detected flux variability, although this is also consistent with the calibration error.
The future simultaneous, multi-frequency ALMA observations at the $>$300 GHz bands, are still required to provide better constraints on the instantaneous spectral energy distributions.
The detailed analysis the Stokes I flux variability, and the analytical modeling of the spectral energy distribution, are beyond the scope of the present paper.
They will be elaborated in our forthcoming papers.

\subsection{Polarization fitting, intrinsic polarization angle, and rotation measure}
\label{sub:pol}
During the observations, the target source is rotated with respected to the receiver frame, due to the alt-azimuth mount of the ALMA antennas. 
If the polarization position angle and the polarization percentage do not vary significantly in a shorter timescale than the durations of our observing tracks (1-2 hours, typically), then the polarization percentage and the polarization position angle can be fitted according to the following formula by definition:
\begin{equation}
\frac{Q}{I} - \delta \equiv \frac{XX- YY}{2I} - \delta = P\cdot\cos(2(\Psi - \eta - \phi) ),
\end{equation}
where $Q$ denotes the observed Stokes Q flux, $\delta$ (Q offset, hereafter) is an assumed constant normalized offset of observed Stokes Q due to amplitude calibration errors or polarization leakage; $P$ is the polarization percentage; $\Psi$, $\eta$, and $\phi$ are the polarization position angle in the sky (e.g. right ascension/declination) frame, the parallactic angle, and the Evector (Section \ref{chap_obs}, Table \ref{table:feed}), respectively (see also Hildebrand et al. 2000; Li et al. 2005).
The application to the interferometric observations of point sources is straightforward since the visibility amplitude does not vary with {\it uv} distance. 
A generalized formulation for extended sources is given in Mart{\'{\i}}-Vidal et al. (2016).

In practice, we determine our fitting results and errors using an iterative process.
First, for each epoch of observations at each frequency, we perform a prior fitting of $P$, $\delta$ and $\Psi$, based on XX and YY observed at various $\eta$. 
We then derive from all data points, the standard deviation ($\sigma^Q$) of the difference between the observed $(XX- YY)/2I - \delta$ and the priori fit.
Finally, we disturb each data point of $(XX- YY)/2I$ with a Gaussian random number of which the standard deviation is $\sigma^Q$, and then re-fit $P$, $\Psi$, and $\delta$.
In the end, $P$, $\Psi$, and $\delta$ and their errors are determined by mean and standard deviations of fittings to 1000 independent realizations of the perturbed $(XX- YY)/2I$.
We have checked that those means converged to the values close to a priori fittings. 
We reject records which present poor signal to noise ratios with the measurements of Stokes I fluxes, which can be observations at low elevation, or with poor weather condition, or are observations of mosaic fields of which the pointing centers are far away from the Sgr A* (e.g. outside of the central 7 fields, see Figure \ref{fig:field12m}, \ref{fig:fieldaca}).

We refer to more details of the observational results of band 8 (on UTC 2015 April 30) to Liu et al. (2016).
Other observational results are introduced in the following sections. 
Our fitting results are summarized in Table \ref{table:pol}.

\begin{figure}
\includegraphics[width=9.5cm]{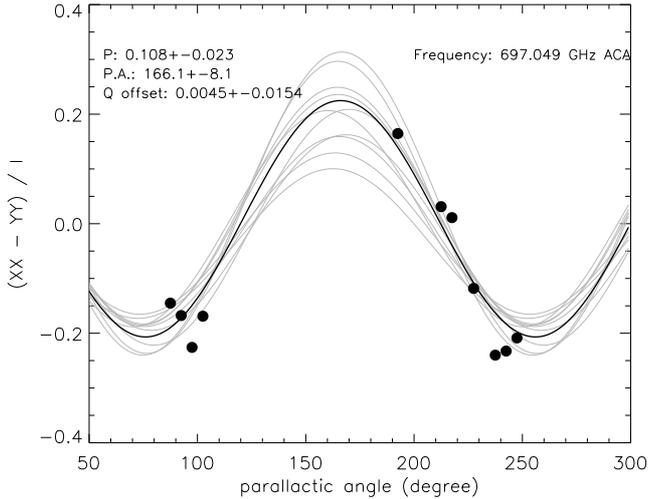}
\caption{Fittings of the (XX-YY)/I intensity ratio of the ACA setups B9-a and B9-b observations together. Symbols, lines, and labels are similar to those in Figure \ref{fig:fit2015b9}. The frequency listed in the upper right is the mean of the observing frequencies of the B9-a and the B9-b setups.}
\label{fig:fit2015b9acacomb}
\end{figure}

\begin{figure}
\vspace{-1cm}
\hspace{-0.25cm}
\begin{tabular}{p{7cm} }
\includegraphics[width=9.5cm]{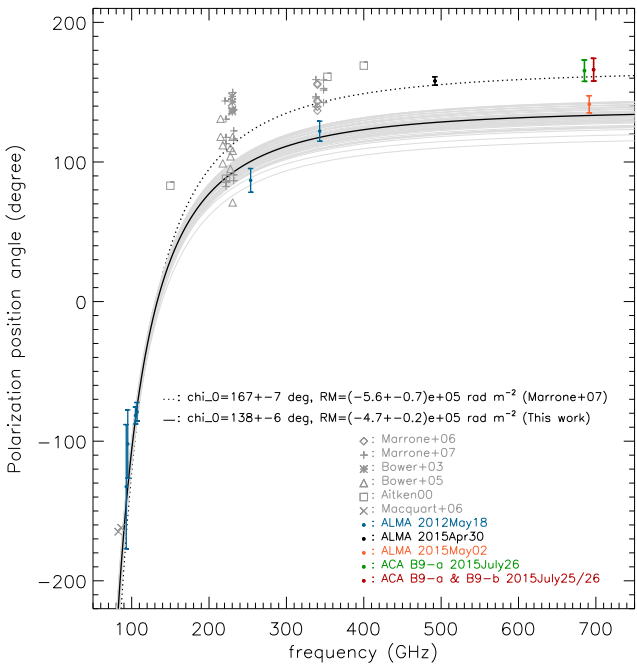} \\
\end{tabular}
\vspace{-1cm}
\caption{\small{The observed polarization position angle at ALMA band 3, 6, 7, 8 and 9 is compared with prior data from Aitken et al. (2000), Bower et al. (2003, 2005), the mean of Macquart et al. (2006), and Marrone et al. (2006a, 2007). The polarization position angles of the Macquart et al. (2006) data were unwrapped by -180$^{\circ}$. Vertical error bars are $\pm$1$\sigma$ uncertainties, which were determined using the procedure introduced in Section \ref{sub:pol}. We overplot the mean fitted intrinsic polarization position angle the and rotation measure by Marrone et al. (2007), and the updated fit based only on ALMA band 3,6,7 data. Gray curves show 50 independent random realizations which characterize our fitting errors.
}}
\label{fig:chiplot}
\vspace{-0.25cm}
\end{figure}

\subsubsection{Band 3 (90-110 GHz), 6 (250 GHz) and 7 (340 GHz) observations on UTC 2012 May 18}\label{subsub:b367}
We fitted $P$, $\Psi$, and $\delta$ for the four spectral windows taken with band 6 and band 7 together, given that the frequency separations of these spectral windows are small as compared with the observing frequencies of these two bands. 
The four spectral windows of band 3 were fitted separately.
The $(XX-YY)/I$ values of the passband averaged band 6 and band 7 data, and those of the four spectral windows of the band 3 data, are plotted in Figure \ref{fig:fit2012may18}.
We additionally provide fittings to the calibrator observations, in Appendix Section \ref{appendix:cal}.

We found that the observed $(XX-YY)/I$ at each frequency can be approximated by assuming constant $P$, $\Psi$, and $\delta$.
However, as compared with the much nicer fits to the calibrators J1924-2914 and NRAO530 (see Appendix), the data of the Sgr A* appear more scattered from the sinusoidal fits, which may be due to small time variation of the polarization position angle and polarization percentage.
There is a monotonic increase of polarization percentage from low to high frequency.
In particular, we found that the two $>$100 GHz spectral windows of band 3 present $\sim$3 times higher polarization percentage than those two $<$100 GHz spectral windows.
Since data of the four spectral windows of band 3 were taken simultaneously, and were calibrated uniformly, their differences in polarization percentage are unlikely to be due to calibration defects. 
The polarized signal is only marginally detected at the two $<$100 GHz spectral windows of band 3, and therefore fittings were subject to large uncertainties.

\subsubsection{Band 9 (680-710 GHz) observations in 2015}\label{subsub:b9}
We fitted $P$, $\Psi$, and $\delta$ for the band 9, 12m-array observations taken on 2015 May 02, and the ACA observations taken on 2015 July 25/26. 
The four spectral windows of these band 9 observations were fitted together, to enhance the signal to noise ratio.
We have checked and confirmed that the $(XX-YY)/I$ values measured from the four spectral windows are consistent, which can be expected by the observed rotation measure of the Sgr A* (see Section \ref{chap_discussion} for more discussion).

The $(XX-YY)/I$ values of the passband averaged band 9 data are plotted in Figure \ref{fig:fit2015b9}.
We found that the observed $(XX-YY)/I$ of the band 9, 12m-Array observations, and the ACA observations of setup B9-a, can be approximated by assuming constant $P$, $\Psi$, and $\delta$.
The good data in both epochs of ACA observations of setup B9-b only covered small parallactic angle ranges, and therefore do not independently provide constraints on $P$ and $\Psi$.
Nevertheless, overplotting the fitting results for the observations of setup B9-a (Figure \ref{fig:fit2015b9}, top right) on the B9-b measurements (Figure \ref{fig:fit2015b9}, bottom row) indicate that the $P$, $\Psi$, and $\delta$ during the time period for the B9-b observations, are consistent with those during the time period of for the B9-a observations.
Fitting together all B9-a and B9-b observations (Figure \ref{fig:fit2015b9acacomb}) therefore may provide a better constraint on $P$ and $\Psi$.

\begin{figure}
\vspace{-0.5cm}
\hspace{-0.75cm}
\begin{tabular}{p{7.5cm} }
\includegraphics[width=10cm]{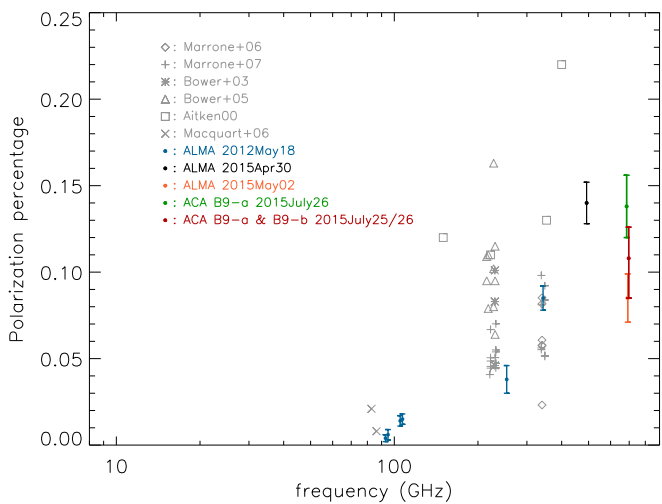} \\
\end{tabular}
\vspace{-0.3cm}
\caption{\small{The observed polarization percentage at ALMA band 3, 6, 7, 8 and 9 is plotted with prior data from Aitken et al. (2000), Bower et al. (2003, 2005), the mean of Macquart et al. (2006), and Marrone et al. (2006a, 2007). Vertical error bars are $\pm$1$\sigma$ uncertainties, which were determined using the procedure introduced in Section \ref{sub:pol}.
}}
\label{fig:fracplot}
\vspace{-0.2cm}
\end{figure}

In spite of the $\sim$3 months of time separations of the band 9 12m-Array and the ACA B9-a observations, and their largely different $\delta$, they present consistent (within 1$\sigma$) polarization position angle $\Psi$.
The measured $P$ from these of band 9 observations are different by a factor of $\sim$1.6.
However, the parallactic angle range covered by these observations (Figure \ref{fig:fit2015b9}) may not permit precisely constraining the polarization percentage $P$, which may tend to be slightly overestimated.
A better constraint on $P$ requires future observations which incorporate polarization calibration, or cover sufficiently large parallactic angle ranges. 
Presently, the $P$ and $\Psi$ constrained by the combined ACA B9-a and B9-b data seem to show deviation of $\Psi$ from the band 9 12m-Array observations (Figure \ref{fig:chiplot}), but show consistent (within 1$\sigma$) $P$ (Figure \ref{fig:fracplot}).
These observed values of $P$ at band 9 ($\sim$700 GHz) are comparable or lower than the value of $P$ observed at band 8 ($\sim$500 GHz).
More discussion of these observations will be defer to Section \ref{chap_discussion}.

\section{Discussion} 
\label{chap_discussion}
Figure \ref{fig:chiplot} and \ref{fig:fracplot} summarize the measured polarization position angles and polarization percentages from our ALMA observations, and compare with the earlier observations of Aitken et al. (2000), Bower et al. (2003, 2005), Macquart et al. (2006), and Marrone et al. (2006a, 2007).
The polarization position angles presented in Figure \ref{fig:chiplot} have taken the polarization feed alignments (i.e. Table \ref{table:feed}) into considerations.
Our observations cannot distinguish the 180$^{\circ}$ ambiguity of the polarization position angle.
The presented polarization position angles in Figure \ref{fig:chiplot} were $\pm$180$^{\circ}$ unwrapped to the nearest possible values to the previous observations.

Least square fitting of the band 3, 6 and 7 observations on 2012 May 18 yield $\chi_{0}$=138$^{\circ}$$\pm$6$^{\circ}$, and RM of (-4.7$\pm$0.2)$\times $10$^{5}$ rad\,m$^{-2}$, where $\chi_{0}$ and RM are the assumed constant intrinsic polarization position angle and rotation measure (e.g. Marrone et al. 2006a).
The RM measured from 2012 May 18 is very close to the previously measured RM=(-5.6$\pm$0.7)$\times $10$^{5}$ rad\,m$^{-2}$ reported by Marrone et al. (2007),and RM=(-4.4$\pm$0.3)$\times $10$^{5}$ rad\,m$^{-2}$ reported by Macquart et al. (2006).
However, Marrone et al. (2007) and Macquart et al. (2006) both reported the derived intrinsic polarization position angles of $\chi_{0}$$\sim$167$\pm$7$^{\circ}$.
The fitted $\chi_{0}$ from the 2012 May 18 observations deviates significantly from those of the earlier observations.
Benefited by the good signal-to-noise ratios provided by the ALMA observations, and the coverage of a large frequency range of the 2012 May 18 observations, we clearly demonstrate that the scattering of the polarization position angles at $\sim$230 GHz and $\sim$345 GHz from the existing measurements (Figure \ref{fig:chiplot}) is at least partially (if not fully) attributed to the time variation of polarization properties instead of merely calibration errors.
The expected polarization position angle at 492 GHz from extrapolating fitting of the 2012 May 18 observations, also is not consistent with the direct measurement made on 2015 April 30.
The polarization position angle measured from the band 9, 12m-array observations (UTC 2015 May 02) is very close to ($\sim$1$\sigma$) the fitting of the band 3, 6 and 7 observations on 2012 May 18 (Figure \ref{fig:chiplot}), however, deviates significantly from the fitting of earlier data (e.g. Marrone et al. 2007). 
The intrinsic polarization position angle of the Sgr A* may be varying/oscillating with time, which requires to be confirmed with observations with better precision.
The polarization position angle measured from the band 9, ACA B9-a observations on 2015 July 26 is consistent with the fitting of earlier data.
The combined ACA B9-a and B9-b observations on 2015 July 25-26 also shows consistent $\Psi$ with the fittings of earlier data (e.g. Marrone et al. 2007).
We note that our analysis here does not consider the effect that the observations at different frequencies may trace the different photospheres.
The better analytic modeling of the polarized emission (e.g. Huang et al. 2009) is still required to fully understand the data presented in this manuscript.

The deviations in Stokes I fluxes (Figure \ref{fig:sed}) and polarization position angles (Figure \ref{fig:chiplot}) of the Band 9 12m-Array observations, from those measured by the ACA B9-a/b observations, are indicative.
However, with two epochs of observations only, our information about whether or not, or how the values of I, $\Psi$ and $P$ at $\sim$700 GHz are correlated, remains limited.
In the case that the polarization percentage indeed remains the same when the total flux increases, then this suggests that the polarized emission may be associated with the flare. 
If the polarization position angle remains constant during the flare, then this suggests that there may be a stable underlying magnetic field which is not perturbed by the energy of the flare. 
A time varying polarization position angle during the flare may on the other hand suggests the growing and collapsing of magnetic loops which have relatively low magnetic field strength.
If future observations showed a smooth evolution of polarization position angle, then propagation of radiating particles around a stable B field geometry is an interesting possibility.
There are more complicated radiative transfer effects which need to be quantified in the ray tracing calculations.
For example, changing in Stokes I flux will also result in changing of the source opacity, which could lead to variations in both polarization position angle and polarization percentage due to depolarization and Faraday rotation within the emission source.
These are beyond the scope of the present paper, which focuses on measurements.

We note that the derived $P$ from the band 9 12m-Array observations, and from the combined band 9 ACA B9-a and B9-b observations, appear lower than the derived $P$ from the band 8 observations (Figure \ref{fig:fracplot}).
The 12m-Array observations of band 8 and 9 were close in time.
The consistent $P$ derived from the band 9 12m-Array observations with the derivation from the combined ACA B9-a and B9-b observations may suggest that $P$ at $\sim$700 GHz does not vary significantly on daily or monthly timescales. 
$P$ and $\Psi$ cannot significantly vary on the hourly (and shorter) timescales, otherwise will prohibit our fittings of polarization using the procedure outlined in the beginning of Section \ref{sub:pol}.
Although our sampling in the time domain remains sparse at the observing frequency of band 9, we consider it is less likely that the lower $P$ at band 9 ($\sim$700 GHz) than that at band 8 ($\sim$500 GHz) is merely due to time variation (Figure \ref{fig:fracplot}).
It is more likely related to the polarization properties of the photospheres probed at these observing frequencies (e.g. Liu et al. 2007).

Finally, the presented new measurements do not yet detect the 90$^{\circ}$ flip of polarization position angle around the transitional observing frequency from the optically thick to the optically thick regimes, as expected by theories (Bromley et al.2001; Liu et al. 2007; Huang et al. 2009).
It may require higher ($>$1 THz) frequency observations.

\begin{table}
\caption{Polarization properties of Sgr A*}
\label{table:pol}
\hspace{0cm}
\begin{tabular}{ l c c c c}\hline\hline
Frequency	& Array & UTC	&  P  & $\Psi^{1}$ \\
(GHz)		&		&		& (\%) & (degree) \\\hline
92.995		& 12m	& 2012May18	& 0.4$\pm$0.2 & -133$\pm$55	 \\
94.932		& 12m	& 2012May18	& 0.6$\pm$0.3 & -102$\pm$49	 \\
104.995		& 12m	& 2012May18	& 1.4$\pm$0.3 & -82$\pm$6	 \\
106.995		& 12m	& 2012May18	& 1.5$\pm$0.3 & -79$\pm$6	  \\
253.750		& 12m	& 2012May18	& 3.8$\pm$0.8 & 87$\pm$5	 \\
342.979		& 12m	& 2012May18	& 8.5$\pm$0.7 & 122$\pm$10  \\
486.150		& 12m	& 2015Apr30	& 14$\pm$0.4 & 158$\pm$3 \\
685.237		& ACA	& 2015Jul26	& 14$\pm$2 & 165$\pm$24 \\
691.537		& 12m	& 2015May02	& 8.5$\pm$1.4 & 141$\pm$14 \\
697.049$^{2}$		& ACA	& 2015Jul25/26 & 11$\pm$2 & 166$\pm$6 \\\hline
\end{tabular}
\vspace{0.2cm}
\footnotesize{ \par $^{1}$ Polarization position angle. \par $^{2}$ Measurements made from combining all ACA observations of B9-a and B9-b setups}
\end{table}

\section{Conclusion} 
\label{chap_conclusion}
We present new measurements of the Stokes I intensity, the polarization position angle, and the polarization percentages for the Galactic supermassive black hole Sgr A*, at frequencies $\sim$100, $\sim$230, $\sim$345, $\sim$500, and $\sim$700 GHz.
We found that the Stokes I intensity at $\sim$700 GHz may be lower than that at $\sim$500 GHz, which suggests that the observations at $\gtrsim$500 GHz may be well into the submillimeter-hump where the emission is becoming optically thin.
At $\sim$700 GHz, both the Stokes I intensity and the polarization position angle may be varying with time, while the observed polarization percentage is consistent with no obvious variations.
After comparing with the previous and the newly reported observations at lower frequency (90-490 GHz), we found that the intrinsic polarization position angle of Sgr A* may be varying with time as well.
Below 500 GHz, we see a monotonic increase of polarization percentage with frequency.
Our observations indicate that the polarization percentage at $\sim$700 GHz may be lower than that at $\sim$500 GHz, which remains to be confirmed with simultaneous measurements at these two frequency bands.

\begin{acknowledgements}
This paper makes use of the following ALMA data: ADS/JAO.ALMA 2013.1.00071.S (PI: Hauyu Baobab Liu), 2013.1.00126.S (PI: Paul T.~P. Ho), and  2011.0.00887.S (PI: Heino Falcke). 
ALMA is a partnership of ESO (representing its member states), NSF (USA) and NINS (Japan), together with NRC (Canada) and NSC and ASIAA (Taiwan), in cooperation with the Republic of Chile. 
The Joint ALMA Observatory is operated by ESO, AUI/NRAO and NAOJ.
We thank Shin'ichiro Asayama, Ted Huang, Hiroshi Nagai and Charles Hull for providing clarification on the ALMA feed orientation.
\end{acknowledgements}

%
%

\appendix

\section{Polarization of calibrator data}\label{appendix:cal}
\begin{figure*}
\begin{tabular}{p{9cm} p{9cm}}
\includegraphics[width=9.5cm]{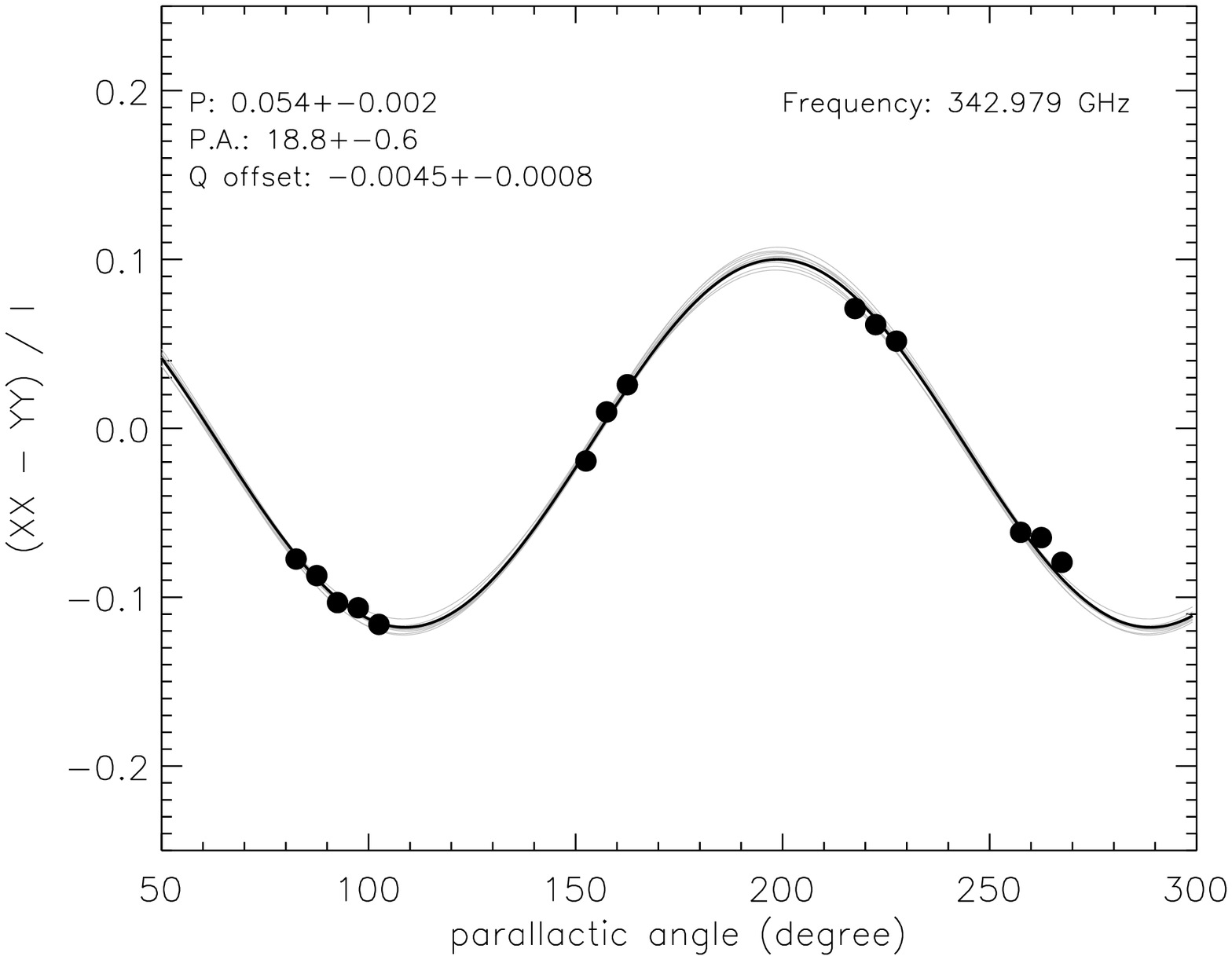} & \includegraphics[width=9.5cm]{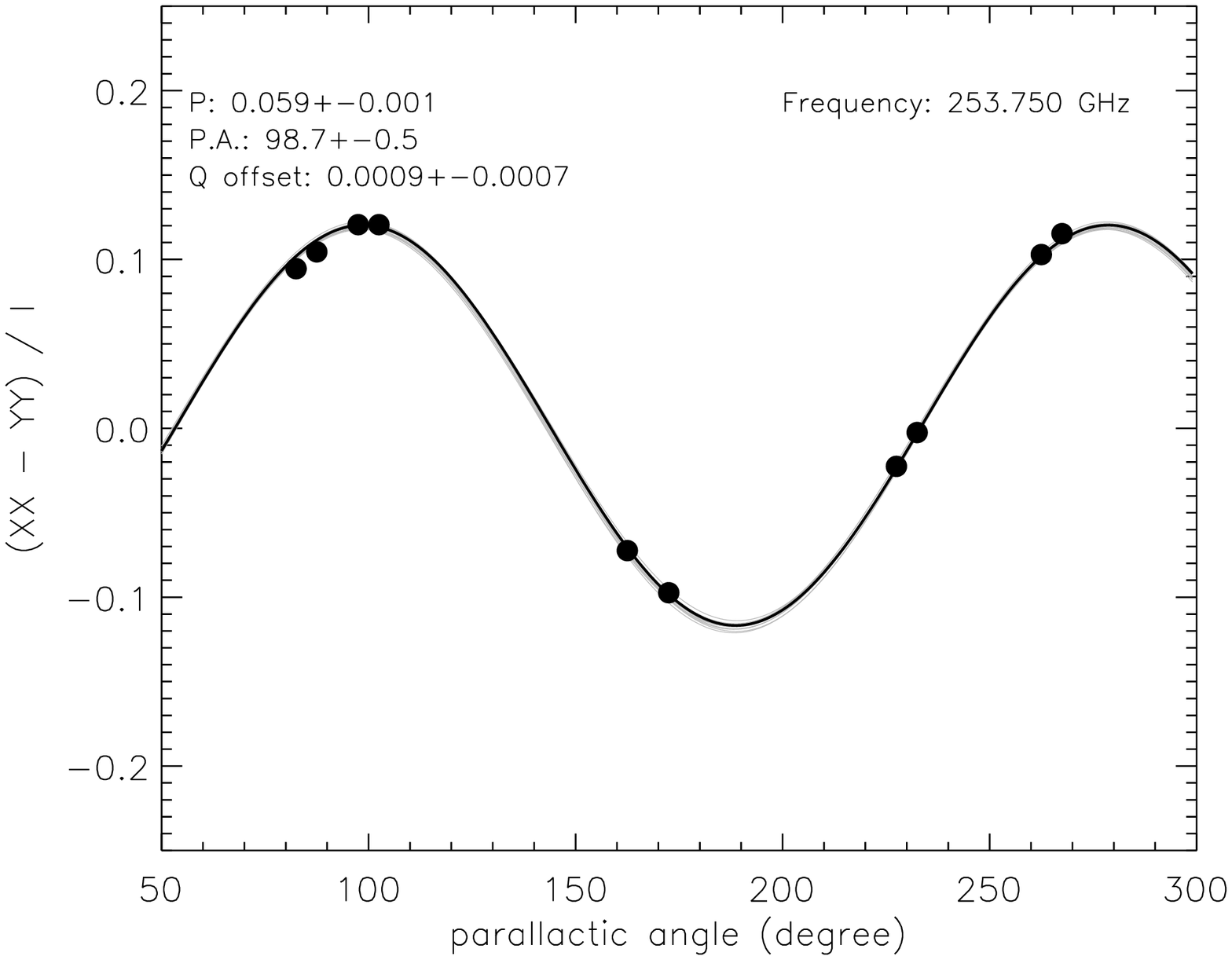} \\
\includegraphics[width=9.5cm]{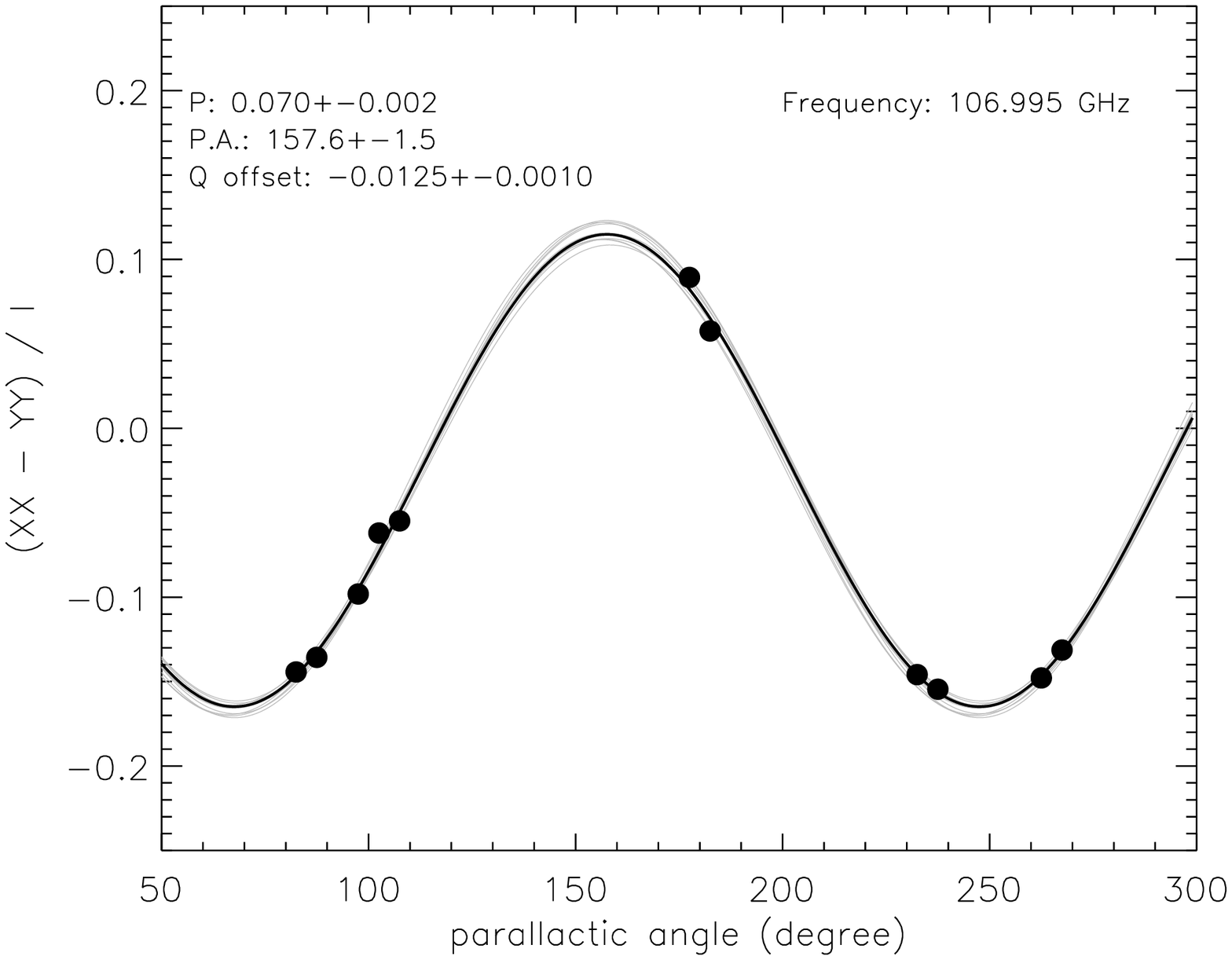} & \includegraphics[width=9.5cm]{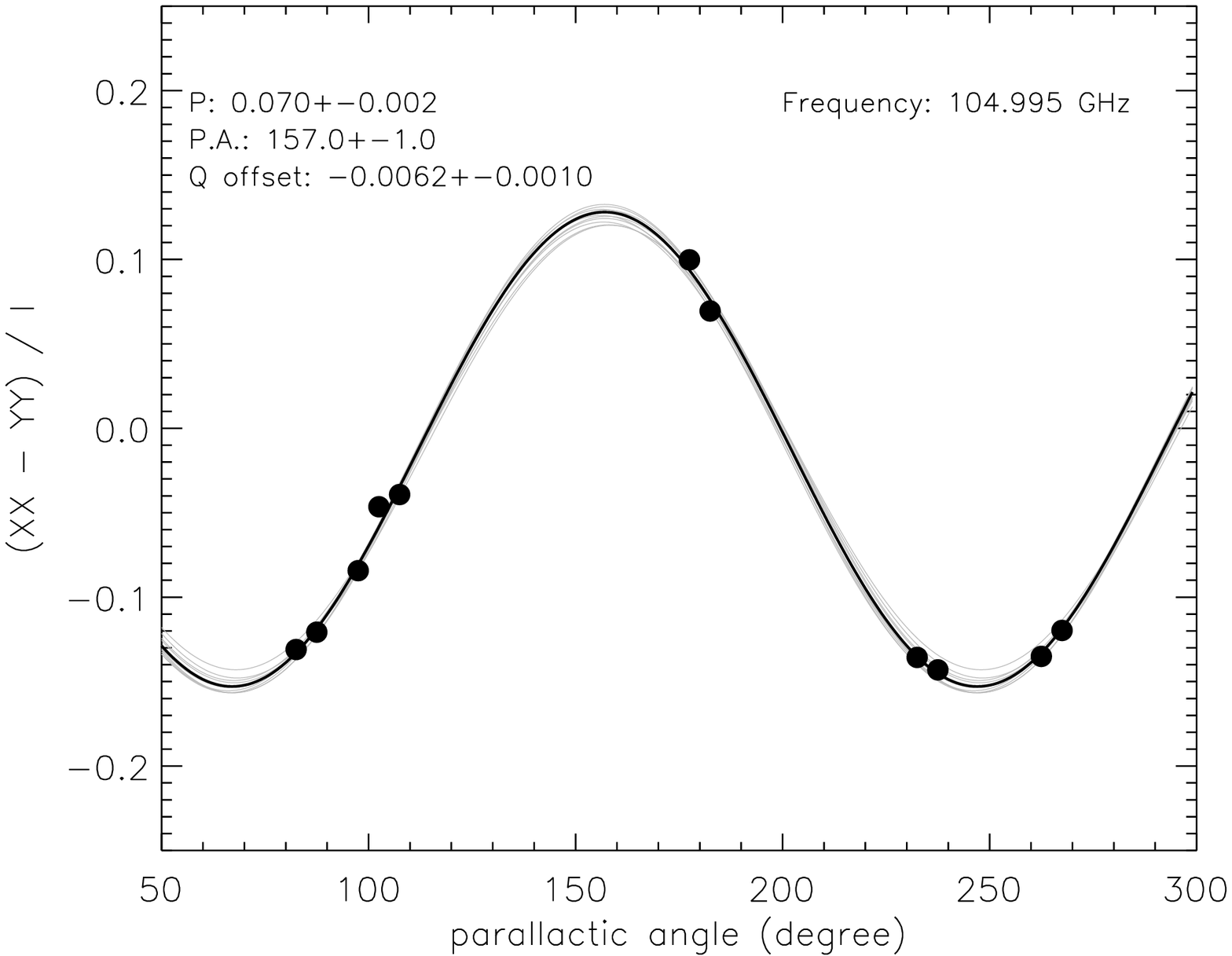} \\
\includegraphics[width=9.5cm]{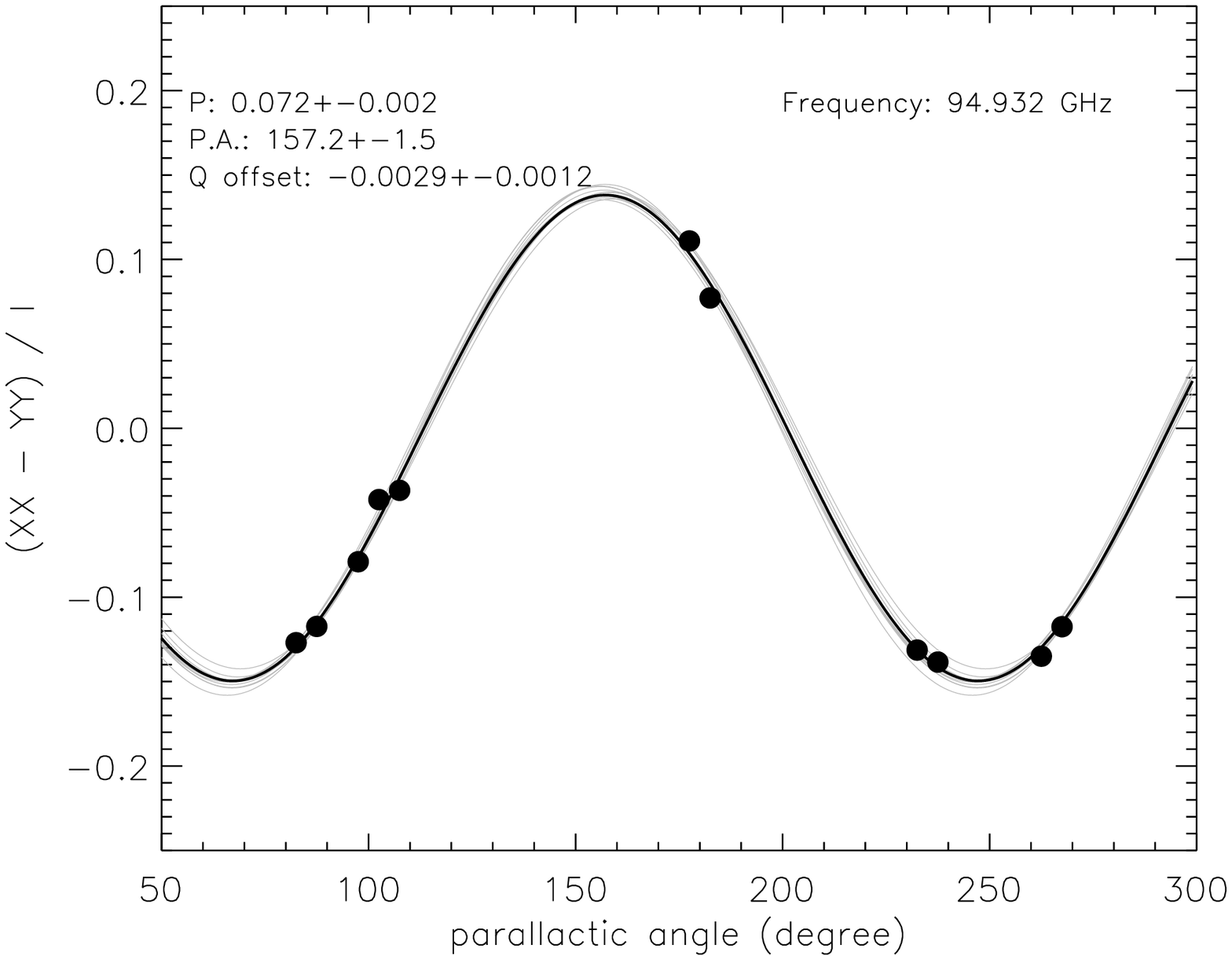} & \includegraphics[width=9.5cm]{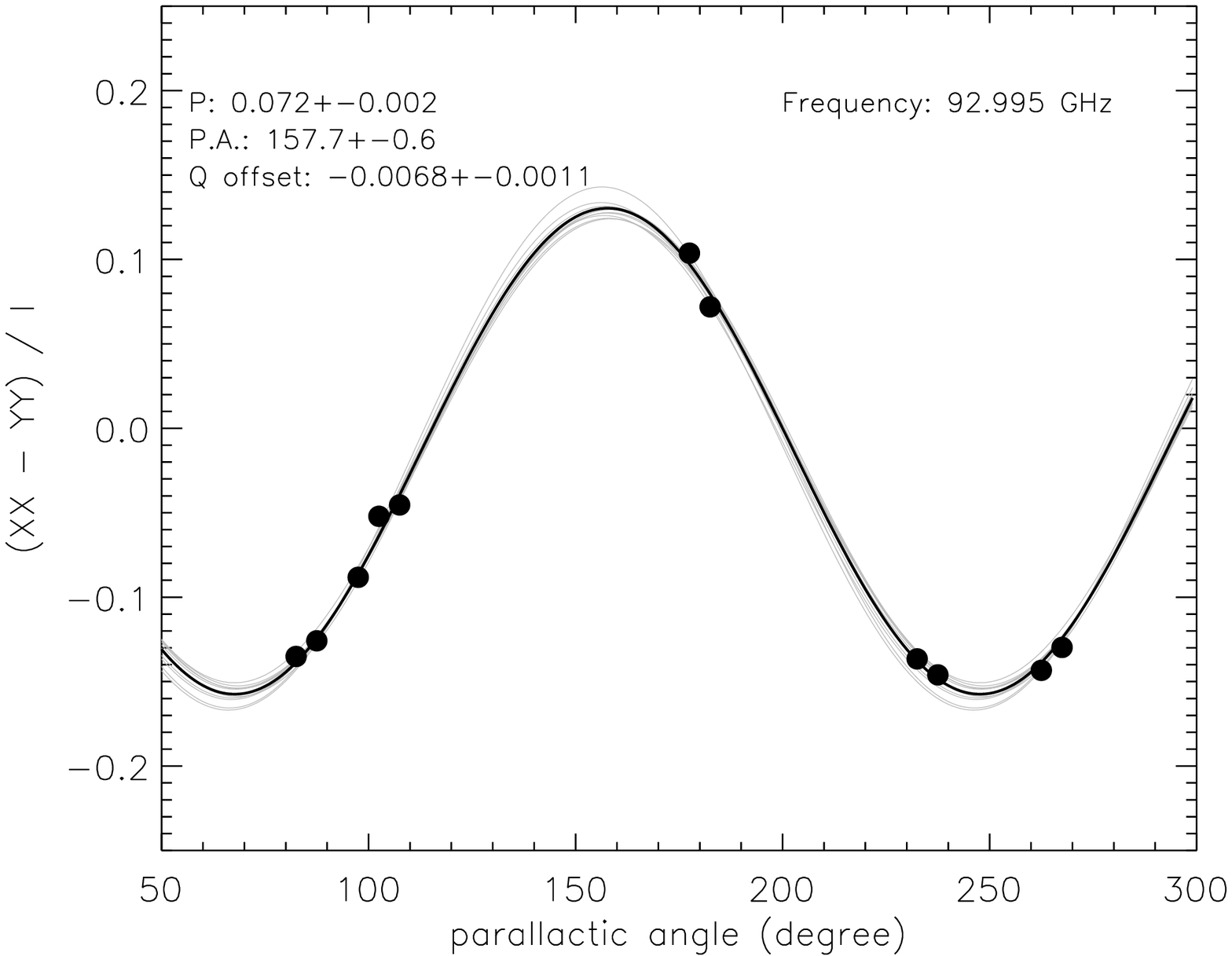} \\
\end{tabular}
\caption{Fittings of the (XX-YY)/I intensity ratio, to determine polarization percentages and polarization position angles. Data presented in this figure are self-calibrated ALMA band 3, 6, 7 observations of J1924-2914 on UTC 2012 May 18. The best fits of polarization percentage (P), polarization position angle (in the receiver frame, i.e. $\Psi-\phi$; P.A.), and a constant normalized Stokes Q offsets (Q offset), are provided in the upper left of each panel, which are represented by a black curve. For each observed frequency, errors of fitted quantities were determined by one standard deviation of fittings of 1000 random realizations of noisy data (details are in Section \ref{sub:pol}). Gray lines in each panel plot every 100 of the random realizations.}
\label{fig:fit2012may18_j1924}
\end{figure*}

\begin{figure*}
\begin{tabular}{p{9cm} p{9cm}}
\includegraphics[width=9.5cm]{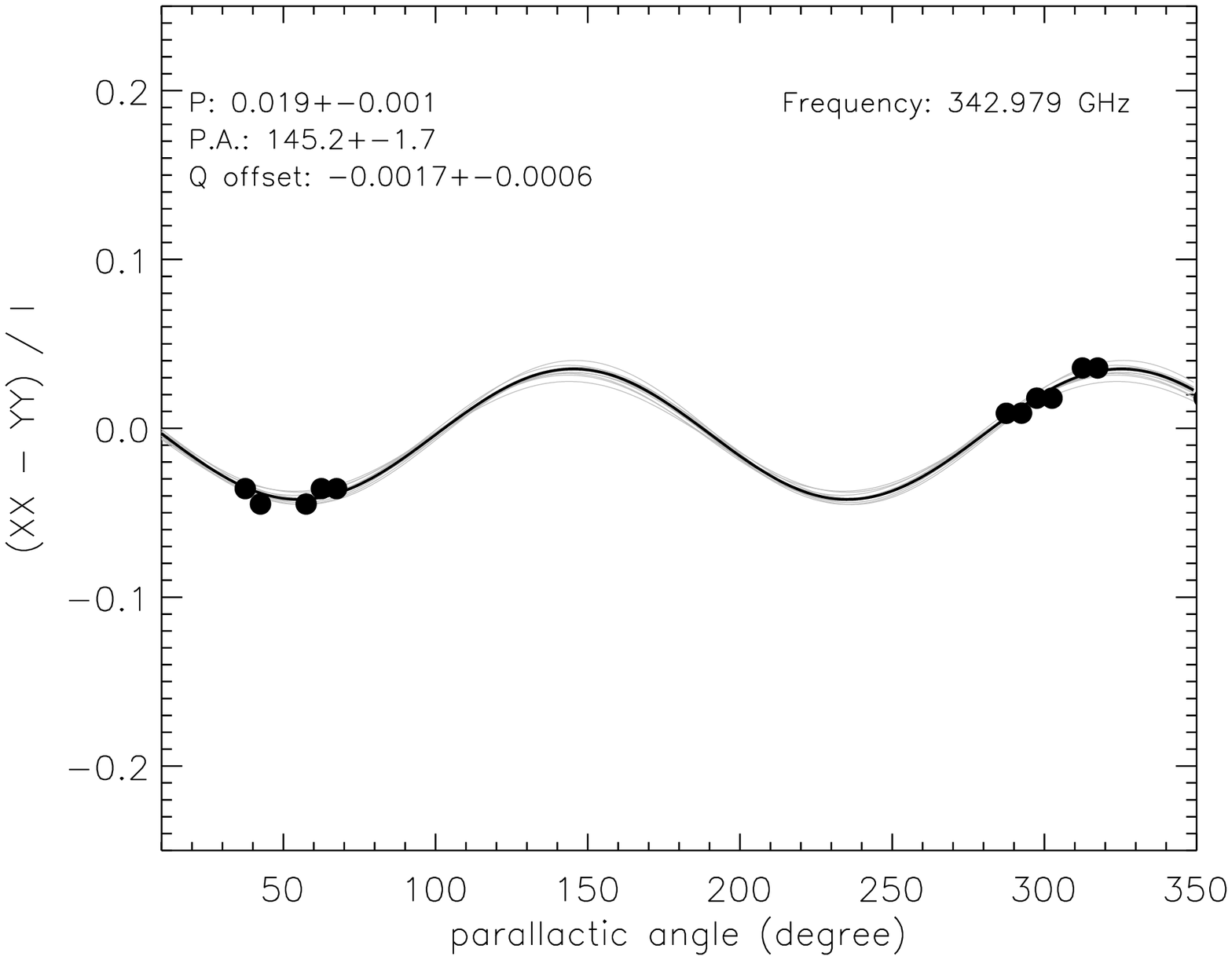} & \includegraphics[width=9.5cm]{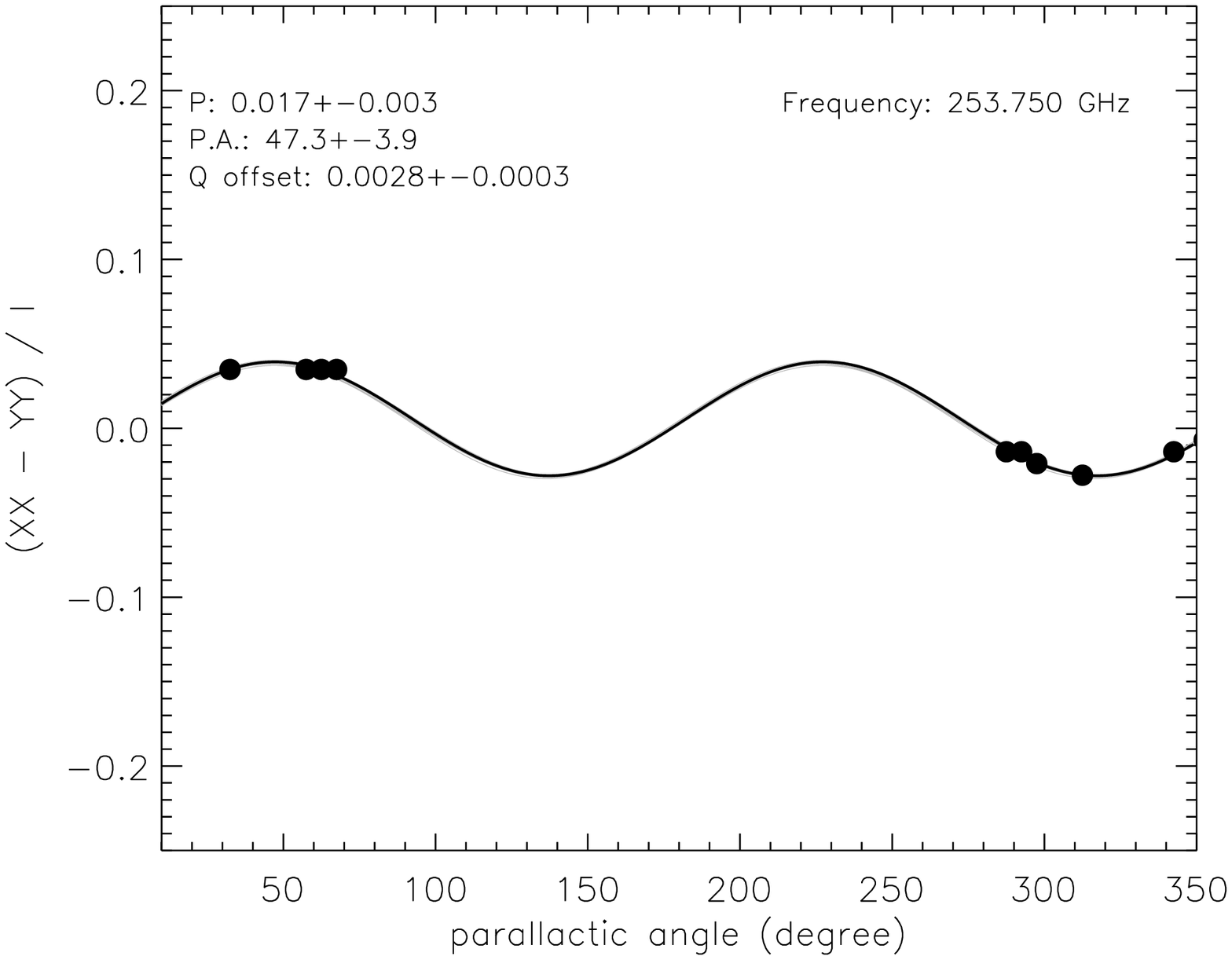} \\
\includegraphics[width=9.5cm]{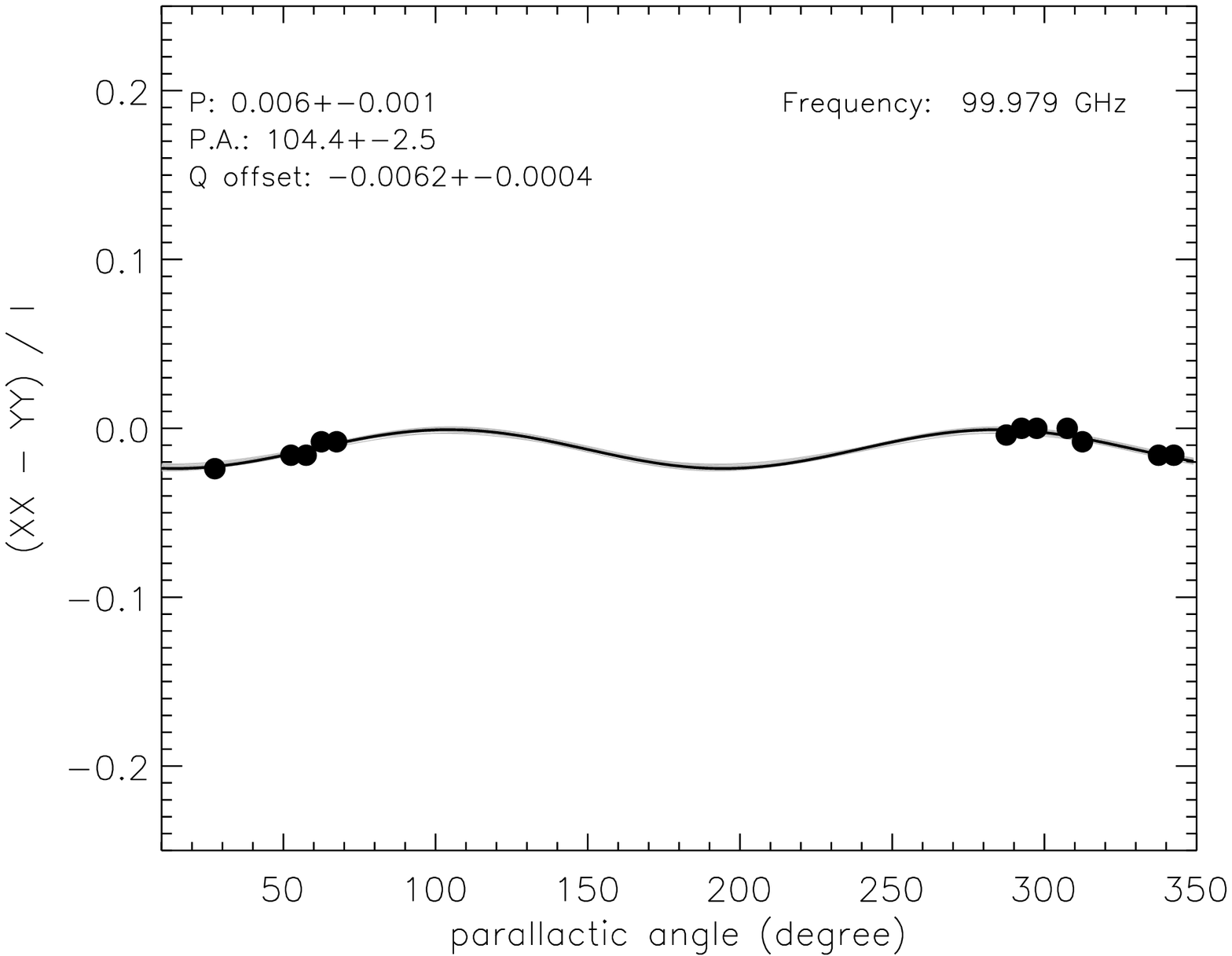} &  \\
\end{tabular}
\caption{Fittings of the (XX-YY)/I intensity ratio, to determine polarization percentages and polarization position angles. Data presented in this figure are self-calibrated ALMA band 3, 6, 7 observations of NRAO530 on UTC 2012 May 18. We fitted the four spectral windows of the band 3 observations together to enhance the signal to noise ratio. The best fits of polarization percentage (P), polarization position angle (in the receiver frame, i.e. $\Psi-\phi$; P.A.), and a constant normalized Stokes Q offsets (Q offset), are provided in the upper left of each panel, which are represented by a black curve. For each observed frequency, errors of fitted quantities were determined by one standard deviation of fittings of 1000 random realizations of noisy data (details are in Section \ref{sub:pol}). Gray lines in each panel plot every 100 of the random realizations.}
\label{fig:fit2012may18_nrao530}
\end{figure*}

In this section, we report fitting results of polarization properties for calibrators observed on UTC 2012 May 18 by the 12m-Array observations (Figure \ref{fig:fit2012may18_j1924}, \ref{fig:fit2012may18_nrao530}).
Polarization of J1924-2914 has been discovered and mentioned in Brinkerink et al. (2015).
The newly reported band 9 observations in the present manuscript covered too small parallactic angle ranges.
In combination with the effects of the potentially relatively large leakages, the band 9 observations cannot provide good fits.
This paper focuses on fitting polarization properties of point sources.
We refer to Mart{\'{\i}}-Vidal et al.(2015) for a method of constraining polarization properties of spatially resolved sources.
Mart{\'{\i}}-Vidal et al.(2016) formulate in detail the problem of the extraction of polarimetry information from dual-polarization observations. 
These authors also discuss the importance of several instrumental effects (like polarization leakage or beam squint) for the special case of the ALMA antennas.

\end{document}